\documentclass[a4paper]{aa}
\usepackage{graphicx}
\usepackage{txfonts}
\usepackage{color}
\usepackage{natbib}

\bibpunct{(}{)}{;}{a}{}{,} 

\begin{document}
\title{Spectral features of Earth-like planets and their detectability at different orbital distances around F, G, and K-type stars}
\author{
P. Hedelt \inst{1,2,3} \and
P. von Paris \inst{2,3,4} \and
M. Godolt \inst{5} \and
S. Gebauer \inst{5} \and
J.~L. Grenfell \inst{5} \and\\
H. Rauer \inst{4,5} \and
F. Schreier \inst{1} \and
F. Selsis \inst{2,3} \and
T. Trautmann \inst{1}}

\institute{
Institut f\"{u}r Methodik der Fernerkundung, Deutsches Zentrum f\"{u}r Luft- und Raumfahrt, Oberpfaffenhofen, D-82234 We{\ss}ling, Germany \and
Univ. Bordeaux, LAB, UMR 5804, F-33270, Floirac, France \and
CNRS, LAB, UMR 5804, F-33270, Floirac, France \and
Institut f\"{u}r Planetenforschung, Deutsches Zentrum f\"{u}r Luft- und Raumfahrt, Rutherfordstr. 2, D-12489 Berlin, Germany \and
Zentrum f\"{u}r Astronomie und Astrophysik, Technische Universit\"{a}t Berlin, Hardenbergstr. 36, D-10623 Berlin, Germany
}
\date{Received ... / Accepted ...}
\abstract
{In recent years, more and more transiting terrestrial extrasolar planets have been found. Spectroscopy already yielded the detection of molecular absorption bands in the atmospheres of Jupiter and Neptune-sized exoplanets. Detecting spectral features in the atmosphere of terrestrial planets is the next great challenge for exoplanet characterization.}
{We investigate the spectral appearance of Earth-like exoplanets in the habitable zone of different main sequence (F, G and K-type) stars at different orbital distances. We furthermore discuss for which of these scenarios biomarker absorption bands and related compounds may be detected during primary or secondary transit with near-future telescopes and instruments.}
{Atmospheric profiles from a 1D cloud-free atmospheric climate-photochemistry model were used to compute primary and secondary eclipse infrared spectra. The spectra are analyzed taking into account different filter bandpasses of two photometric instruments planned to be mounted to the James Webb Space Telescope (JWST). We analyze in which filters and for which scenarios molecular absorption bands are detectable when using the space-borne JWST or the ground-based telescope E-ELT (European Extremely Large Telescope).}
{Absorption bands of carbon dioxide (CO$_2$), water (H$_2$O) , methane (CH$_4$) and ozone (O$_3$) are clearly visible in both high-resolution spectra as well as in the filters of photometric instruments. However, only during primary eclipse absorption bands of CO$_2$, H$_2$O and O$_3$ are detectable for all scenarios when using photometric instruments and an E-ELT-like telescope setup. CH$_4$ is only detectable at the outer HZ of the K-type star since here the atmospheric modeling results in very high abundances. Since the detectable CO$_2$ and H$_2$O absorption bands overlap, separate bands need to be observed to prove their existence in the planetary atmosphere. In order to detect H$_2$O in a separate band, a S/N of S/N$>7$ needs to be achieved for E-ELT observations, e.g. by co-adding at least 10 transit observations. Using a space-borne telescope like the JWST enables the detection of CO$_2$ at 4.3\,$\mu$m, which is not possible for ground-based observations due to the Earth's atmospheric absorption. Hence combining observations of space-borne and ground-based telescopes might allow to detect the presence of the biomarker molecule O$_3$ and the related compounds H$_2$O and CO$_2$ in a planetary atmosphere. Other absorption bands using the JWST can only be detected for much higher S/Ns, which is not achievable by just co-adding transit observations since this would be far beyond the planned mission time of JWST.}
{}
\titlerunning{The spectral appearance of Earth-like planets}
\maketitle

\section{Introduction}
Spectroscopic or photometric techniques enable the detection and even the investigation of the atmosphere of a transiting extrasolar planet (exoplanet). Increasing instrumental sensitivities allow for the detection of low-mass planets. Furthermore, long-term planet search programmes are able to detect planets that are farther away from their host stars. Potentially rocky planets with masses lower than Neptune, which orbit within the Habitable Zone (HZ) of their central stars, are of main interest. Several potentially terrestrial planets orbiting in or close to the HZ are already known (Kepler-22\,b: \citealp{Borucki2012}, HD 85512 b: \citealp{Pepe2011}, Gliese 581 c and d: \citealt{Udry2007}, \citealt{Mayor2009}, Gliese667C\,c: \citealp{Bonfils2012,Anglada-Escude2012}, HD 40307\,g: \citealp{Tuomi2013}). The search for more of these terrestrial planets and their characterization will eventually address the question, whether life on Earth is unique.

Analyzing the wavelength-dependent extinction of the stellar light passing through the planetary atmosphere during the primary transit allows for the chemical characterization of the planetary atmosphere. For hot-Jupiter planets, the detection of infrared molecular absorption bands of water (H$_2$O) and methane (CH$_4$), as well as of atomic lines in the visible using this technique have already been announced (see e.g. \citealt{Charbonneau2002,Vidal-Madjar2003,Vidal-Madjar2004,Knutson2007,Tinetti2007,Swain2008,Snellen2010}).

Up to now only two super-Earth planets (i.e. planets having masses below 10\,M$_\mathrm{Earth}$) have been investigated in transmission (GJ\,1214\,b, \citealt{Charbonneau2009} and CoRoT-7\,b, \citealt{Leger2009}). \citet{Bean2010} obtained transmission spectra of GJ\,1214\,b (6.55\,M$_\mathrm{Earth}$), which were lacking any spectral features. The nature of its atmosphere is currently under discussion (see e.g. \citealt{Miller-Ricci2012,Bean2011,Croll2011,Desert2011,Crossfield2011,deMooij2012,Berta2012}). For CoRoT-7 b (6.9\,M$_\mathrm{Earth}$, \citealt{Hatzes2010}), \citet{Guenther2011} were able to determine upper limits of the extension of its exosphere by observing the strength of different emission lines using high-resolution transmission spectroscopy.\\

From secondary eclipse observations, i.e. when the planet passes behind its star, both emission and reflection spectra can be obtained. Characteristics of the thermal emission of the planet can be investigated in the infrared, since here the planet-star flux ratio is orders of magnitudes higher than at optical wavelengths. The emergent spectrum holds information about the temperature structure of the atmosphere as well as the atmospheric components. The thermal emission of several hot Jupiter and Neptune-sized exoplanets has already been observed (see e.g. \citealt{Snellen2010b, Richardson2007,Grillmair2007,Deming2006,Alonso2009,Alonso2009b,Sing2009,Demory2012}).

In the optical regime, the stellar light that is reflected from the planet before it moves behind its host star may be observed and thus the planetary albedo could be determined (see e.g. \citealt{Seager2008}). With reflection spectra the existence of clouds can be inferred and constraints on the energy budget can be made (see e.g. \citealt{Rowe2008,Cowan2011}).\\

In order to determine whether a terrestrial planet is habitable or even inhabited, the detection of biomarker molecules plays a key role. These molecules are closely connected to or required for the existence of life as we know it. Biomarker molecules include nitrous oxide (N$_2$O) and  ozone (O$_3$). N$_2$O on Earth has an almost exclusively biogenic origin from nitrifying and denitrifying bacteria \citep{Oonk1998}. O$_3$ indicates the presence of oxygen in the atmosphere, which on Earth is almost exclusively produced by photosynthesis of plants and cyanobacteria. \citet{Selsis2002} and \citet{Segura2007} have discussed a possible abiotic formation of O$_3$ in CO$_2$ dominated atmospheres, whereas \citet{Domagal2010} discussed a possible abiotic O$_3$ buildup for Earth-like planets orbiting M-dwarfs.

Other related compounds connected to habitability and life are H$_2$O, CH$_4$ and carbon dioxide (CO$_2$). Liquid H$_2$O is necessary for life as we know it on Earth. It provides the majority of the greenhouse effect on modern Earth, warming the surface above the freezing point of water. CH$_4$ has both biotic and abiotic sources and is also a strong greenhouse gas. CO$_2$ is important for habitability mostly because of its greenhouse effect (especially towards the outer HZ) and the carbonate-silicate cycle (e.g. \citealt{Walker1981}).\\

Earth is so far the only known example of a habitable and inhabited planet, that can be used to investigate the parameter space of habitable conditions. It is straightforward (at least conceptually) to build a spectrum from a given arbitrary atmospheric composition. The inverse problem, however, i.e. to infer the characteristics of a planet from a spectrum, is much more difficult due to its ill-posed nature. It is nevertheless of paramount importance to understand what type of planet lies behind a given observed spectrum.

A huge number of parameters affect the atmospheric and spectral appearance of a given Earth-like planet. In this paper we investigate the influence of the central star type and the orbital distance to the central star within the HZ. We consider only small variations away from Earth, where our model assumptions (Earth development, Earth biomass, etc.) are more likely to be valid.  We furthermore discuss for which conditions (in terms of orbital distance in the HZ and central star) molecular absorption bands can be detected with near-future telescope facilities.\\

Previous major modeling efforts which calculated synthetic spectra of hypothetical terrestrial extrasolar planets were performed, e.g. by \citet{DesMarais2002}, \citet{Segura2003}, \citet{Segura2005}, \citet{Ehrenreich2006}, \citet{Tinetti2006}, \citet{Kaltenegger2007}, \citet{KalteneggerTraub2009}, and \citet{Rauer2011} to examine the influence of, e.g. different host stars, atmospheric abundances, and atmospheric evolution on the spectral appearance. \citet{Arnold2002}, \citet{Arnold2008}, \citet{Palle2009}, \citet{Vidal-Madjar2010}, \citet{Palle2011}, \citet{Hedelt2011} and \citet{Ehrenreich2012} analyzed observations of the terrestrial planets of the Solar System (i.e. Venus, Mars, and Earth) as proxies for exoplanets.

\citet{Segura2003} presented emission spectra for an Earth-like planet around F, G and K-type stars and discussed the spectral response due to varying O$_2$ concentration. In this work, we vary the orbital distance to study the impact on emission and transmission spectra. In addition to calculating the spectral response, we furthermore discuss the detectability of the spectral features. For this purpose, we assume two telescope configurations and different instrument specifications. Improving upon the approach of \citet{KalteneggerTraub2009} and \citet{Rauer2011}, we present background-limited signal-to-noise ratios (S/Ns) instead of photon-limited S/Ns for a ground-based and a space-borne telescope setup.

\citet{Grenfell2007} investigated the effect on the atmospheric chemistry for slightly different scenarios than the ones considered in this work. Hereafter this paper will be referred to as ``G07''. We use an updated version of their atmospheric model and a slightly different position for the center HZ runs. In this paper we investigate if the chemical responses found lead to detectable spectral signatures.\\

Section \ref{Models} summarizes the atmospheric model, the radiative transfer model as well as the S/N model. Furthermore in this section the telescope and instrument parameters are shown as well as the considered model scenarios. The atmospheric and spectral response are discussed in Sect. \ref{Analysis}, followed by a discussion about the detectability of molecular absorption lines. Section \ref{Discussion} discusses the results, before Sect. \ref{Summary} presents our conclusions.

\section{Methods \& Scenarios}\label{Models}
\subsection{Models}
\subsubsection{Atmospheric model} \label{Atmmodel}
For this paper, we use the one-dimensional, cloud-free coupled climate and photochemical model of \citet{Rauer2011}, which calculates global, diurnally-averaged atmospheric temperature, pressure and concentration profiles. It is based on the model used by G07, including improvements in the coupling of climate and photochemical modules as well as code optimization. A detailed model description can be found in \citet{Rauer2011}, as well as in \citet{Segura2003}, and \citet{Grenfell2007a}. The code optimization includes an improved calculation of the climate module grid and a different convergence criterion (see \citealt{Rauer2011} for details).

Since the model does not incorporate clouds, the model surface albedo (i.e. the reflectivity of the surface) is adjusted until the temperature profile of modern Earth is reproduced when simulating modern-day Earth conditions. In G07, a surface albedo of 0.218 was assumed, whereas for the profiles used in this paper the required value is 0.207 due to code optimization. Although the temperature profile is successfully reproduced, radiative fluxes (e.g. spectral albedos) calculated in the model do not reproduce modern Earth values since clouds are neglected, as stated above. We note that the influence of clouds on spectral albedos has been investigated by e.g. \citet{Robinson2011}, and \citet{Kitzmann2011}.

\subsubsection{Radiative transfer model}\label{RTmodel}
The spherical line-by-line radiative transfer model MIRART-SQuIRRL \citep{Schreier2001,Schreier2003} has been used to calculate high-resolution synthetic emission and transmission spectra. It was designed for the analysis of terrestrial atmospheric measurements and has been verified by intercomparisons with other radiative transfer codes (see e.g. \citealt{Clarmann2003} and \citealt{Melsheimer2005}). MIRART-SQuIRRL uses HITRAN2008 \citep{Rothman2009} for the calculation of absorption cross-sections and the temperature, pressure, water vapor and concentration profiles of 15 species from the atmospheric model, including H$_2$O, CO$_2$, CH$_4$, O$_3$, CO, and N$_2$O. Furthermore continuum absorption corrections for H$_2$O and CO$_2$ are performed.

Emission spectra are calculated from the planetary surface up to the top of the atmosphere (located at $6.6\times10^{-5}$\,bar). In order to obtain a disk-integrated emission spectrum of the planet, we integrate a set of pencil beams with zenith angles from 0$\degr$ (zenith) to 90$\degr$ (horizon) in steps of 1$\degr$. Note that this approach is different to previous studies by, e.g. \citet{Rauer2011} or \citet{Segura2003}, where only one single spectrum has been calculated at a viewing zenith angle of 38$\degr$ and then multiplied by $\pi$ in order to obtain a disk-integrated spectrum. Fully disk-integrated spectra take into account the atmospheric limb darkening. The difference between both approaches is mostly smaller than 10\%, hence does not greatly influence the discussion presented hereafter. We furthermore take into account reflected stellar light that would be measured during secondary eclipse observations. The reflected component is in our scenarios important up to 4\,$\mu$m. We added the contribution of reflected stellar light to the secondary eclipse spectrum by multiplying the stellar spectra with the spectral albedo of the atmosphere, which is calculated by the climate part of our atmospheric model. From these spectra brightness temperature spectra are calculated by identifying spectral fluxes with a blackbody temperature, assuming that the distance and radius of the planets are precisely known. This temperature is related to the physical temperatures of the contributing radiating atmospheric levels. We note that brightness temperature spectra calculated from secondary eclipse spectra yield temperatures which are in the near-infrared much warmer than the planetary temperatures since the spectra additionally include the reflected stellar component. We furthermore note that it might be possible to disentangle the planetary emission spectrum from the reflected component if the stellar spectrum is known. However, this will be challenging due to stellar variability and the low S/Ns found in the near-IR (see Sect. \ref{SNR}).

Transmission spectra ($\mathcal{T}_i(\lambda)$) for 62 adjacent tangential beams through the atmosphere are calculated. The tangential heights $h_i$ correspond to the layer altitudes of the model atmosphere. The diameter of each beam is given by $\Delta h_i=h_\mathrm{max}/n_\mathrm{layers}$, with $h_\mathrm{max}$ the altitude of the model lid (dependent on the scenario considered) and $n_\mathrm{layers}=64$, the number of layers in the photochemical model. The overall transmission $\mathcal T(\lambda)$ is simply the arithmetic mean of all $\mathcal{T}_i(\lambda)$, since beams cross the atmosphere in equidistant grid points. The transit depth $d_\mathrm{Transit}(\lambda)$ for all scenarios considered is calculated by:
\begin{equation}
d_\mathrm{Transit}(\lambda)=\frac{(r_p+h(\lambda))^2}{r_s^2},
\end{equation}
with $r_p$ the planetary radius and $r_s$ the stellar radius. $h(\lambda)$ is the effective height of the atmosphere at a given wavelength $\lambda$:
\begin{equation}
h(\lambda)=\sum\limits_i\left(1-\mathcal{T}_i(\lambda)\right)\Delta h_i.
\end{equation}
The effective height of the atmosphere is the additional obscuring radius provided by the atmosphere that effectively increases the observed size of the planet during the primary transit.

We note that a possible night-side pollution of a transit depth measurement, that was found by \citet{Kipping2010} to be significant for hot-Jupiter planets, can be neglected in our scenarios. The contrast of the planetary to the stellar emission is of the order of $\sim10^{-5}$, whereas for hot-Jupiters it is of the order of $\sim10^{-3}$ \citep{Kipping2010}. The night-side pollution is an effect arising from the normalization of the transit depth (measured from the different fluxes before and during the transit) to the out-of-transit flux.

\subsubsection{S/N model}
We will discuss the detectability of spectral features in Sect. \ref{SNR} assuming the planets are observed by a ground-based or a space-borne telescope. For this we calculate background-limited S/Ns using the code of \citet{vonParis2011}, which is based on the code of \citet{Rauer2011}. Note that this reproduces the values found by \citet{KalteneggerTraub2009}, when using similar assumptions regarding spectral resolution and atmospheric profiles. We improved the code of \citet{vonParis2011} to calculate background-limited S/Ns for ground-based measurements by taking into account the thermal background as well as the transmission of Earth's atmosphere.

To take into account the Earth's atmospheric emission and transmission for ground-based measurements, we use spectra\footnote{available online via \newline http://www.eso.org/sci/facilities/eelt/science/drm/tech\_data/data} provided by ESO \citep{ELT-DRM2010} for the E-ELT (European Extremely Large Telescope) exposure time calculator \citep{Liske2008}. The spectra are provided for different telescope sites and have been calculated using a tropical atmospheric profile and the HITRAN2004 database \citep{Rothman2005}. The water vapor profile has been adjusted to fit the mean precipitable water column value of the telescope site. For the ground-based telescope configuration we use the files for the ``Paranal site'' with an airmass of 1.0 (zenith viewing), a site altitude of 2,600\,m (743\,mbar), an ambient temperature of 285\,K, and a precipitable water column value of 2.3\,mm. The E-ELT will be located at Cerro Armazones at an altitude of 3\,064\,km. However, Earth atmospheric spectra and site parameters are not yet available. Calculated S/Ns are thus likely to be underestimated when using the Paranal site parameters.

For the calculation of S/Ns for ground-based observations, the simulated planetary spectrum is then multiplied with the Earth's transmission spectrum. The atmospheric noise $\sigma_\mathrm{A}$ from the thermal emission of the Earth's atmosphere is treated in the same way as the zodiacal noise (cf. Eq. A.7 in \citealt{vonParis2011}) using the Earth's emission spectrum as input. The telescope emission is modeled as a grey body (i.e. a black body multiplied by a constant emissivity).

We furthermore updated the code of \citet{vonParis2011} in order to calculate S/Ns for photometric instruments, which provide a different detector response. Instead of a refractive element for spectrometric instruments, photometric instruments are using transmissive windows to filter the incoming radiation in certain wavelength bandpasses. Hence the pixel area occupied on the detector is different than that for a dispersed spectrum. The number of pixels occupied on the detector $n_\mathrm{px}(\lambda)$ is calculated from the angular diameter of the Airy disk $\theta$ and the pixel scale $p_s$
\begin{equation}\label{Pix_spatial}
n_\mathrm{px}(\lambda)=\left(\frac{\theta}{p_s}\right)^2=\left(2.4\cdot\frac{\lambda}{D\cdot p_s}\right)^2.
\end{equation}

\subsection{Telescope \& instrument configurations}
In this paper we investigate the spectral response of several absorption bands measured by high-resolution spectroscopic instruments, as well as by photometric instruments. Photometric instruments provide the ability to obtain high S/Ns by integrating the light over a fixed spectral bandpass. We are using the filter bandpasses of MIRI (Mid-InfraRed Instrument, \citealt{Wright2004}) and NIRCam (Near-InfraRed Camera), that are instruments planned for the space-borne James Webb Space Telescope (JWST). MIRI provides 9 filters in the range from 5 to 27.5\,$\mu$m, whereas NIRCam provides filters in the range from 0.6 to 5.0\,$\mu$m. Table \ref{Ins} shows the wavelength bandpasses of all filters and their respective names that are used within this work. Note that we are not using all available filters, but only a selection of filters that are covering strong absorption bands (indicated in the table) or atmospheric windows (indicated as (R) in the table).

\begin{table}[t]
  \centering
  \caption{Operating wavelengths of photometric instrument filters used in this work.}\label{Ins}
  \begin{tabular}{lccc}
    \hline\hline
    Filter & $\lambda$ [$\mu$m] & $\Delta\lambda$ [$\mu$m] & Absorbers\\\hline
    \multicolumn{4}{c}{NIRCam}\\\hline
    F277W & 2.77 &  0.7& H$_2$O, CO$_2$ \\
    F356W (R) & 3.56 &  0.89 & (CH$_4$)\\
    F430M & 4.35 & 0.2 & CO$_2$\\
    F480M & 4.80 & 0.4 & O$_3$\\
    \multicolumn{4}{l}{ }\\
    \multicolumn{4}{c}{MIRI}\\\hline
    IM01 & 5.60 & 1.2 & H$_2$O\\
    IM02 & 7.70 & 2.2 & H$_2$O, CH$_4$\\
    IM03 & 10.00 & 2.0 & O$_3$\\
    IM04 (R) & 11.30 & 0.7 & (H$_2$O) \\
    IM06 & 15.00 & 3.0 & CO$_2$\\
    IM07 & 18.00 & 3.0 & H$_2$O\\
   \hline
  \end{tabular}
  \tablefoot{The table shows the filter names used throughout this work, their center wavelength $\lambda$ and bandpass $\Delta\lambda$, as well as the strongest species absorbing in the filter for the NIRCam instrument (top panel) and the MIRI (bottom panel), respectively. Reference filters covering atmospheric windows are indicated by (R).}
\end{table}

We will furthermore discuss the detectability of molecular absorption bands, when using the JWST as an example of a space-borne telescope and the E-ELT of a ground-based telescope. JWST is scheduled for launch in 2018 and will provide a telescope aperture of 6.5\,m diameter. The E-ELT will be located in the Atacama desert in Chile, on Mountain Cerro Armazones (3\,064\,m altitude) and is planned to be operational by 2020. The E-ELT will have a primary mirror of 39.3\,m diameter and a central obstruction of 11.76\,m diameter. Ground-based telescopes observing in the infrared are restricted to certain atmospheric windows, where the Earth's atmosphere has a high transmissivity.

The E-ELT will be equipped with EPICS (Exo-Planet Imaging Camera and Spectrograph) which will be designed for the near-IR (0.6 - 1.65\,$\mu$m) and with METIS (Mid-infrared E-ELT Imager and Spectrograph) which will perform spectro-photometry in the wavelength range from 3 to 14\,$\mu$m. Since the instrumentation is still in the planning phase and the filters are not yet defined, we use the wavelength bandpasses of the NIRCam filters for EPICS and that of MIRI for METIS, in order to allow for a comparison of JWST and E-ELT capabilities. The pixel scale $p_s$ and dark current $d_c$ for both instruments that are needed for the calculation of background-limited S/Ns are however already selected and will be used in our calculations. Table \ref{Params} lists the telescope and instrumental parameters that are used for the S/N calculations.

\begin{table}[t]
  \centering
  \caption{Telescope and instrument parameters used for S/N calculations.}\label{Params}
  \begin{tabular}{lcc}
    \hline\hline
    Parameter & JWST$^{(1)}$ & E-ELT$^{(2)}$ \\\hline
    Aperture $d_t$ [m]& 6.5 & 39.3\\
    Detection efficiency $q$  & 0.15 & 0.5\\
    Emitting area $A_\mathrm{Telescope}$ [m$^2$]& 240 & 1104.42 \\
    Temperature $T_\mathrm{Telescope}$ [K] & 45 & 285 \\
    Emissivity $\epsilon$ & 0.15 & 0.14\\\hline
  \end{tabular}

  \begin{tabular}{lcccc}
  \\
    \hline\hline
    Parameter & NIRCam$^{(1)}$& MIRI$^{(1)}$&  EPICS$^{(3)}$ & METIS$^{(4)}$\\\hline
    $p_s$ [mas\,px$^{-1}$]& 65  & 110 & 2.3 & 17.2\\
    $d_c$ [e$^-$ px$^{-1}$] & 0.01 & 0.03 & 0.5 & 2000 \\
    \hline
  \end{tabular}
  \tablefoot{The top panel shows the telescope parameters used in this paper, whereas the second panel shows the instrument parameters.}
  \tablebib{
  (1) ~\citet{KalteneggerTraub2009,Deming2009,Belu2011};
  (2) \citet{ELT-DRM2010};
  (3) \citet{Kasper2010};
  (4) \citet{Kendrew2010}.
  }

\end{table}

\subsection{Model scenarios}\label{Scenarios}
\begin{table}[t]
  \centering
  \caption{Scenarios considered in this paper. }\label{ScenarioTable}
  \begin{tabular}{ccccc}\hline\hline
  Position & Orbital distance & $T_\mathrm{Surf}$ & $F_\mathrm{ToA}$&$t_\mathrm{transit}$\\
  \qquad & [AU] & [K] & [W/m$^2$]&[h]\\\hline
  \multicolumn{5}{c}{Planet around F2V star ($r_s=1.06\times R_{\sun}$ $^{(1)}$)}\\\hline
  Inner run & 1.706& 303.3 & 1642.38 & 13.79\\
  Center run & 1.888 & 280.2 & 1337.44 & 14.50\\
  Outer run & 1.961 & 273.0 & 1242.48 & 14.77\\
  \multicolumn{5}{l}{ }\\
  \multicolumn{5}{c}{Planet around G2V star ($r_s= R_{\sun} = 695\,500$\,km)}\\\hline
  Inner run & 0.936 & 303.2 & 1549.88& 12.58\\
  Center run &1.000 & 288.1 & 1357.76 & 12.98\\
  Outer run & 1.079 & 273.0 & 1165.80 &13.48\\
  \multicolumn{5}{l}{ }\\
  \multicolumn{5}{c}{Planet around K2V star ($r_s=0.81\times R_{\sun}$ $^{(2)}$)}\\\hline
  Inner run & 0.582 & 303.2 & 1465.93 & 8.68\\
  Center run & 0.606 & 294.3 & 1353.59 & 8.83\\
  Outer run & 0.675 & 273.0 & 1089.64 &9.33\\\hline
  \end{tabular}
  \tablefoot{The three panels show for each central star chosen (F-type star: top, G-type star: center, K-type star: bottom) the orbital distance, surface temperature $T_\mathrm{Surf}$, incident top-of-atmosphere stellar flux $F_\mathrm{ToA}$ and transit time $t_\mathrm{Transit}$ for each scenario. Transit times have been calculated assuming an inclination of 90$\degr$.}
  \tablebib{(1)~ \citet{Habing2001}; (2) \citet{Santos2004} }
\end{table}

We model the atmosphere of an Earth-sized planet with a planetary radius of $r_p=1$R$_{\oplus}$ and a mass of $m=1$M$_{\oplus}$. We employed for each stellar type the same initial Earth-like composition as G07 (1\,bar surface pressure and, e.g. 21\% O$_2$, 355\,ppm CO$_2$, 1\% Ar and N$_2$ as filling gas) as well as Earth's biomass surface emissions (prescribed as in \citealt{Rauer2011}). The sample stars used for the calculation of atmospheric profiles are the Sun as a well-known G2V star, $\sigma$ Bootis as a representative of an F2V star and $\epsilon$ Eridani for the K2V star. Note that throughout the paper the different central star types used are referred to as G-type star, F-type star and K-type star, respectively.

The solar spectrum is constructed from high-resolution observations \citep{Gueymard2004} and the $\sigma$ Bootis spectrum is taken from \citet{Segura2003}. Compared to G07 and \citet{Segura2003}, we use a different stellar spectrum for $\epsilon$ Eridani (see \citealt{Kitzmann2010}). We furthermore use a different stellar distance for the F-type reference star $\sigma$ Bootis. Using the Hipparcos parallax of 64.66\,mas for $\sigma$ Bootis, we obtain a distance of 15.5\,pc, which is confirmed by \citet{Habing2001}. This is substantially larger than the 12\,pc used by G07 and \citet{Segura2003}.

The planets are positioned at orbital distances to their central star, such that for the ``inner runs'' (the model runs with the planetary orbital distance closest to their star) the model surface temperature reaches $\sim$303\,K (30$^{\circ}$C). The outer edge was defined with a surface temperature of $\sim$273\,K (0$^{\circ}$C).  These limits corresponds to a definition of the HZ given by \cite{Dole1964} for complex life. Note that modern day HZ definitions allow a much wider range, but it is challenging for this model setup to calculate the full range, since the photochemistry and climate calculations can only be carried out within a narrow temperature range. Classical studies of the runaway greenhouse effect at the inner edge of the HZ (e.g. \citealt{Kasting1988,Kasting1993}), calculate a troposphere which is fully saturated with water vapor. These hot and humid conditions are difficult to study with our coupled climate and chemistry model. For the more-limited range of temperatures that we consider, we assume an Earth-like relative humidity profile from \citet{Manabe1967} (see also G07). For the outer HZ the formation of clouds and their climatic impact as well as the carbonate-silicate cycle would become very important. This cycle controls the CO$_2$ content of the atmosphere on the Earth such that it increases with decreasing surface temperature, thus stabilizing the climate and extending the outer HZ. This means that our outer runs are unlikely to represent fully self-consistent scenarios, since we keep the CO$_2$ concentration constant over the entire range of orbital distances considered. However, such consistent coupling between atmospheric and geochemical processes is difficult and beyond the scope of this paper.

For the central runs the modeled planets are positioned at an orbital distance from their central star where the stellar energy input equals that of the present-day mean total solar irradiance ($F_\mathrm{ToA}=1\,366$\,W\,m$^{-2}$), as in \citet{Kitzmann2010}, and \citet{Rauer2011}. This approach was chosen so that habitable surface temperatures were not enforced and which allowed the temperature to adjust in a consistent way to the different spectral energy distributions provided by the central stars. In this way the effects arising from the different spectral energy distributions are separated from variations of the total incoming flux. Note that in previous studies (e.g. \citealt{Segura2003,Segura2005}, and G07), the center of the HZ was defined such that the calculated surface temperature reached 288\,K. However, the qualitative chemical responses on varying the orbital distance in this work are mostly similar to G07, although the absolute values change.

Table \ref{ScenarioTable} lists all scenarios considered with estimated transit times $t_\mathrm{transit}$ and the top-of-atmosphere (ToA) wavelength-integrated fluxes $F_\mathrm{ToA}$. Stellar radii needed for the calculation of transit depths are also given. The transit times are given for one single transit either during primary or secondary eclipse. Compared to G07, we need to apply different orbital positions for the planets orbiting F and K-type stars. For the F-type star this is related to the change in the normalization of the observed stellar spectrum, whereas for the K-type star the stellar input spectrum was differently constructed than that in G07 (see \citealt{Kitzmann2011}).

\section{Results}\label{Analysis}
In this section we will first shortly present the atmospheric response on varying the orbital distance. We will show the spectral response in high-resolution spectra before discussing what can be measured by the different filters provided by the instruments discussed in this paper, and if this would be detectable.

\begin{figure}[t]
\centering
  \resizebox{\hsize}{!}{\includegraphics*{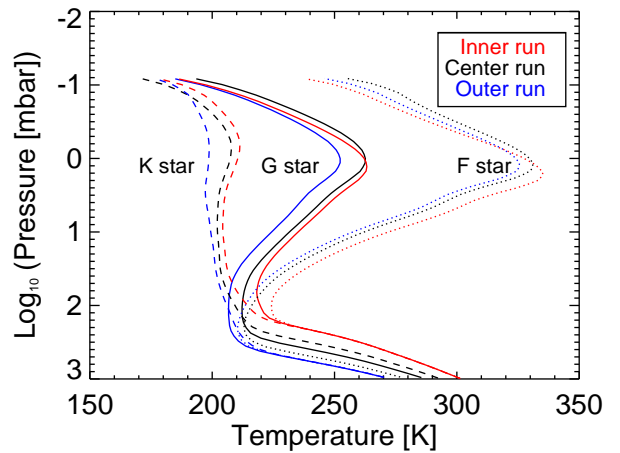}}
  \caption{Pressure-temperature profiles of the Earth-like planets around different central stars (K-type star: dashed, G-type star: solid, F-type star: dotted) at different orbital distances (inner run: red, center run: black, outer run: blue).}\label{Temps}
\end{figure}

\subsection{Atmospheric response}\label{AtmosResponse}
On increasing the orbital distance, the variation of the temperature profile is much smaller than for changing the central star type (see Fig. \ref{Temps}). Note that we find surface temperatures of about 300\,K at an orbital distance of 0.94\,AU for a planet orbiting a Sun-like star, whereas other authors find temperatures above $\sim350$\,K (see e.g. \citealt{Kasting1993,Selsis2007}). Our atmospheric model uses a fixed relative humidity profile of \citet{Manabe1967} (see G07). Earlier calculations regarding the inner boundary of the HZ \citep{Kasting1988,Kasting1993} assumed instead a saturated troposphere, i.e. a relative humidity of unity. Using an isoprofile would result in higher temperatures by about 30\,K, even for present solar insolation. We neglect the feedback between surface temperature and relative humidity, hence we are likely underestimating surface temperatures and water abundances for the inner runs.

\begin{figure}[t]
\centering
\resizebox{\hsize}{!}{\includegraphics*{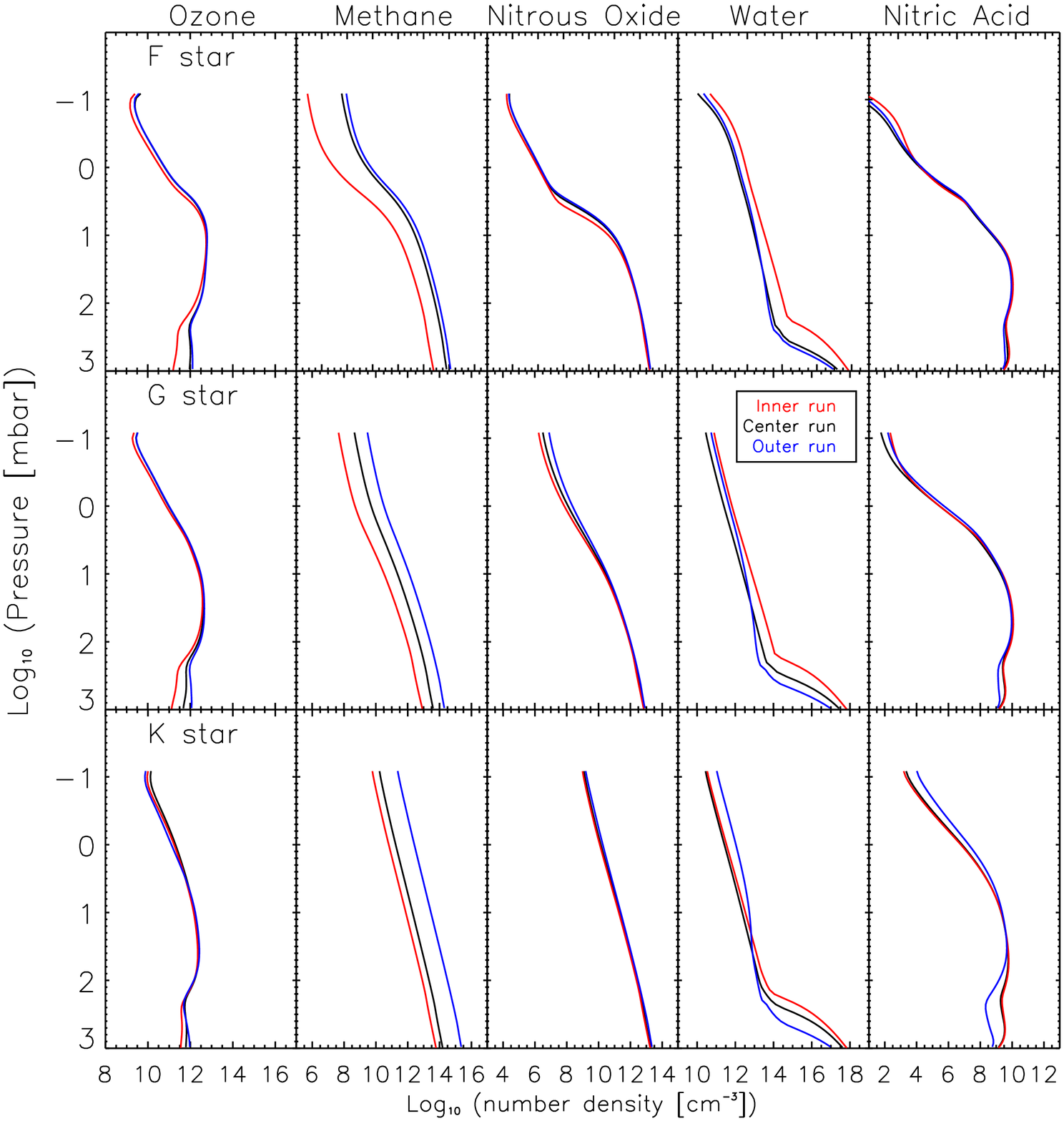}}
  \caption{Pressure-concentration profiles of five prominent spectral species of the Earth-like planets around different central stars (upper row: F-type star, center row: G-type star, lower row: K-type star). Columns from left to right show profiles of O$_3$, CH$_4$, N$_2$O, H$_2$O and HNO$_3$. The orbital distance is indicated by colors as in Fig. \ref{Temps}. Note that CO$_2$ is set to an isoprofile of 355\,ppm and hence not shown.} \label{Bios}
  \end{figure}
\begin{figure}
\centering
\resizebox{\hsize}{!}{\includegraphics*{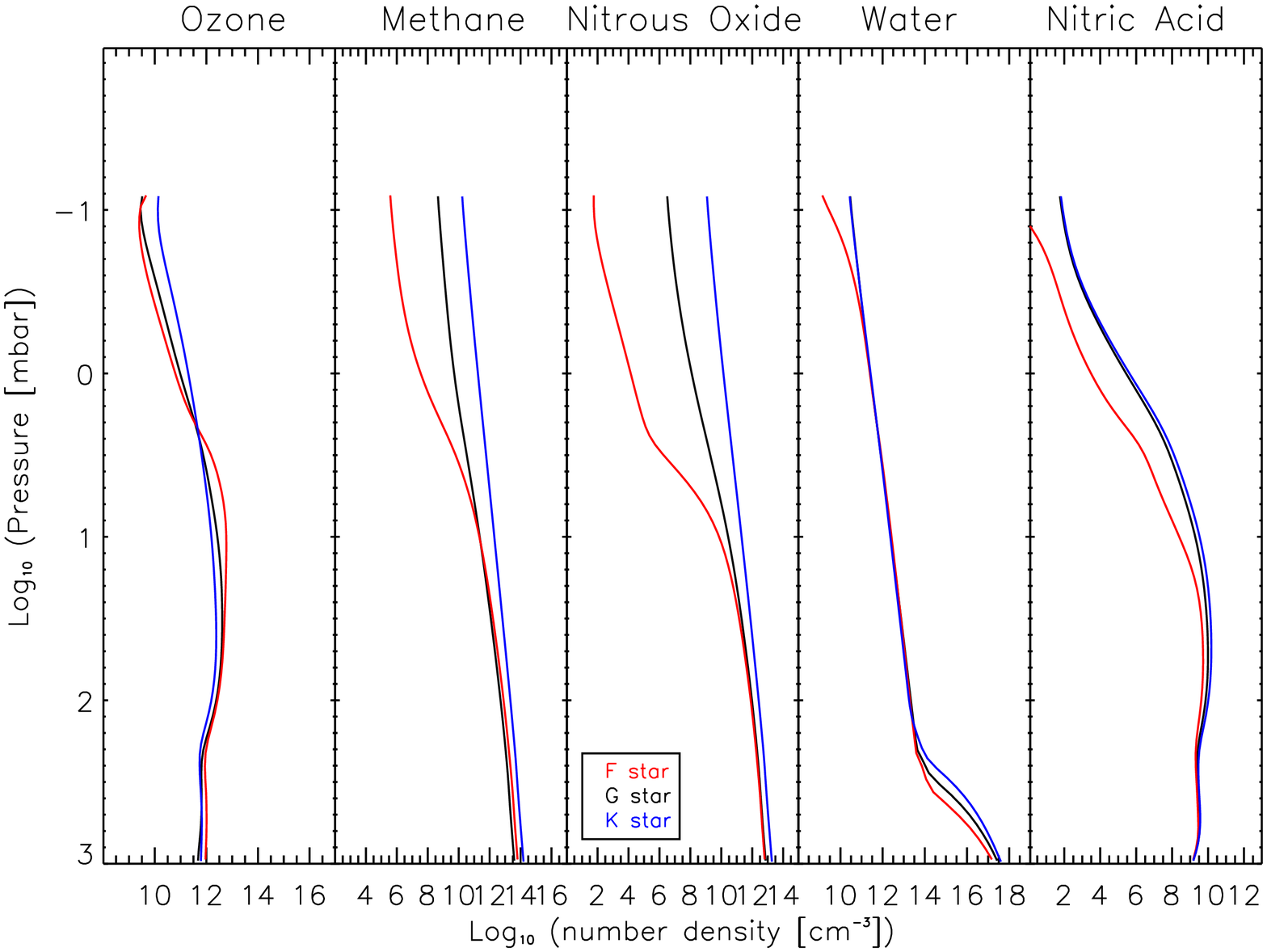}}
\caption{Same as Fig. \ref{Bios}, but only the profiles of the planets at the center of the HZ are shown to visualize the dependency on central star type. The central star type used is color-coded (red: F-type star, black: G-type star and blue: K-type star)}.  \label{BiosStars}
\end{figure}

The orbital distance variation has the strongest effect on H$_2$O and CH$_4$. Figs. \ref{Bios} and \ref{BiosStars} show the corresponding profiles of five prominent spectral species, whereas Table \ref{DUs} shows the change in the total column amounts when increasing the orbital distance from the inner to the outer boundary. In comparison to G07 we find a somewhat stronger increase for O$_3$ and CH$_4$ and a smaller increase for N$_2$O. H$_2$O is controlled by the tropospheric water content, which is identical by construction. Nevertheless, the conclusions presented in G07 do not change when using the updated model.

\begin{table}[t]
  \centering
  \caption{Relative change of total column amount when increasing the orbital distances from the inner to the outer boundary for different molecules and central star types.}\label{DUs}
  \begin{tabular}{lccc}
  \hline\hline
    Molecule  & F-type&G-type&K-type \\\hline
O$_3$&1.37&1.46 (1.10)&0.94\\
CH$_4$&22.09 & 24.24 (17.36)& 37.04\\
N$_2$O &  1.11 &  1.12 (1.23)&  1.24\\
H$_2$O &  0.08 &  0.08 (0.08)&  0.08\\
HNO$_3$ &  0.71 &  0.73 (-)&  0.56\\\hline
  \end{tabular}
    \tablefoot{Values given in brackets are the values found by G07.}
\end{table}

\subsection{Spectral response}
Transmission spectroscopy during primary eclipse probes the molecular composition of the atmosphere. Furthermore, the height of the atmosphere which is related to the pressure and atmospheric structure can be determined, since the atmosphere becomes optically thick at different altitudes for different wavelengths.

\begin{figure*}[ht]
\centering
\includegraphics[width=8.5cm]{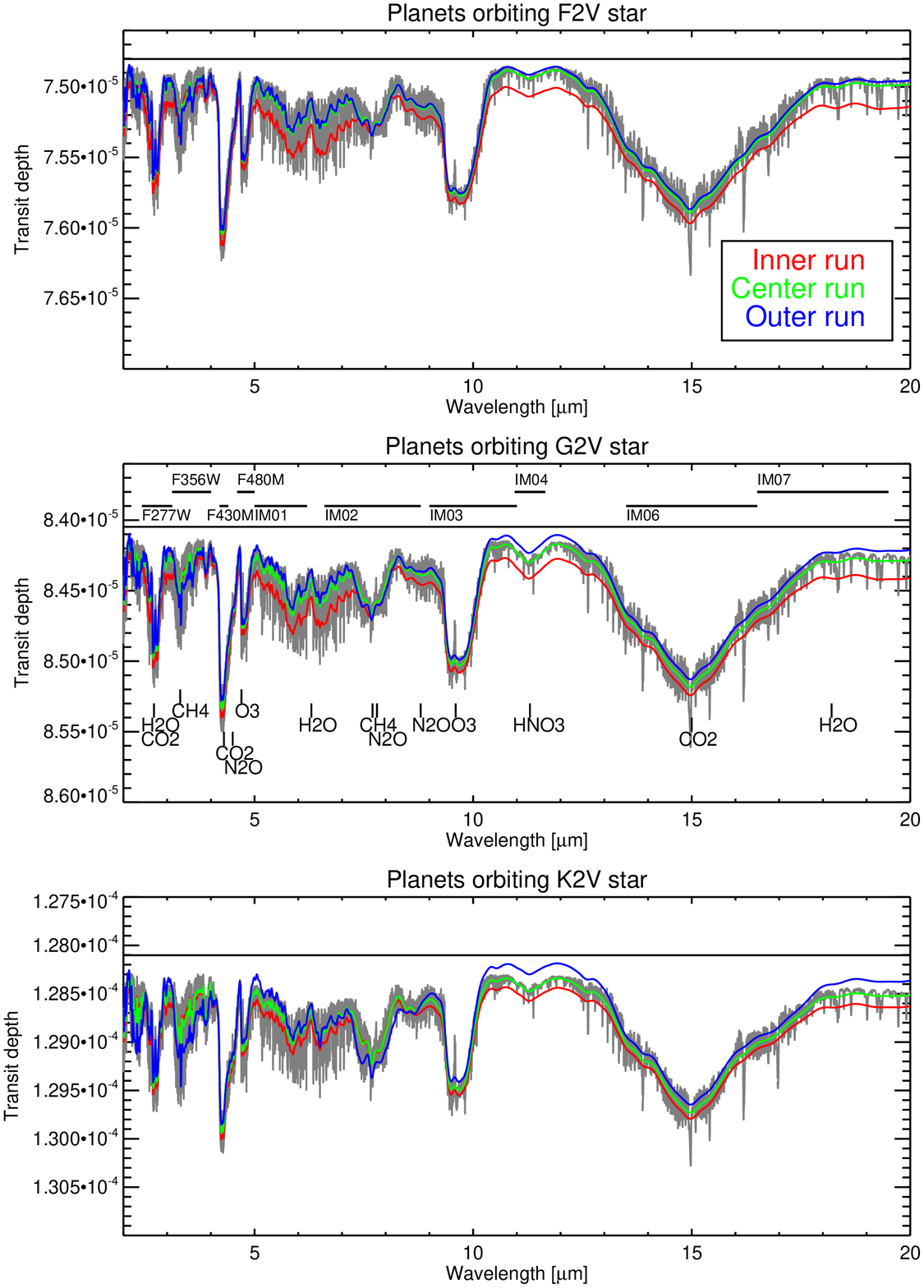}
\includegraphics[width=8.5cm]{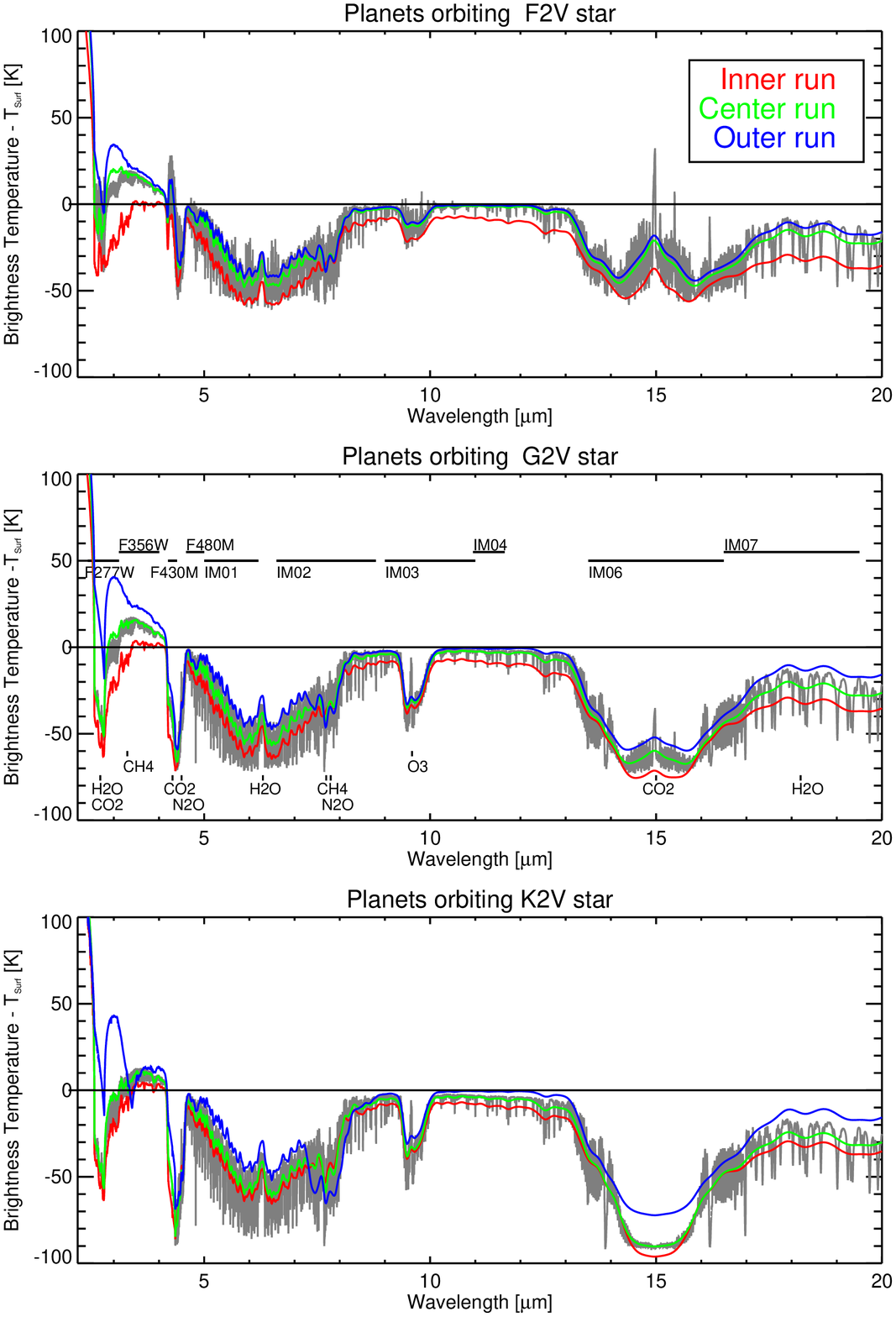}
\caption{Transit depth during primary eclipse (left) and brightness temperature difference with respect to the calculated surface temperature spectrum during secondary eclipse (right) for the scenarios considered. The spectral resolution is $R=100$. Each center run with $R=3\,000$ is shown in grey. The geometric transit depth (see Sect. \ref{FilterResponse}) is indicated by a horizontal line for transmission spectra. The brightness temperature spectra include the reflected stellar component in the near-IR. Furthermore the bandpass of the filters considered in this work are shown.}\label{HighResSpecs}
\end{figure*}

Observation of the secondary eclipse provides contrast spectra of the planetary and stellar emission. In the near infrared up to 4\,$\mu$m the reflection of the stellar spectrum dominates the secondary eclipse spectrum, whereas at higher wavelengths the planetary emission dominates.
A brightness temperature spectrum computed from secondary eclipse radiance spectra shows the range of temperatures that can be found in a given atmosphere. We however note that to obtain a brightness temperature spectrum from a measured radiance spectrum, the planetary radius and the distance of the planet to the observer need to be well-known, which we assume to be the case here. The brightness temperature spectrum yields temperatures high above the planetary temperatures in the near-infrared due to the reflected stellar light (see Fig. \ref{HighResSpecs}, right). Thus the stellar and planetary fractions may be separated. We note that the spectral albedo is dominated by absorption rather than scattering, since Rayleigh scattering is negligible in the wavelength range from 2 to 4\,$\mu$m and the main species responsible for Rayleigh scattering (N$_2$, O$_2$, CO$_2$) are constant for all scenarios considered. With increasing orbital distance, the spectral albedo increases due to less H$_2$O absorption (except for the outer K star run, where the CH$_4$ absorption increases strongly). This increase in spectral albedo for the F and G star runs is approximately of the same amount as the stellar irradiation decreases (see Table \ref{ScenarioTable}). Thus the reflected component changes only slightly with increasing orbital distance. Note that Fig. \ref{HighResSpecs} shows the brightness temperature difference to the model surface temperature. Since the surface temperature decreases by 30\,K (by construction) and the reflected component changes only slightly, the reflected component seems to increase with increasing orbital distance.\\

Figure \ref{HighResSpecs} shows for two different spectral resolutions the transit depth for primary eclipse transmission spectra and the brightness temperature difference to the model surface temperature for the secondary eclipse. Both transit depth and brightness temperature spectra clearly show strong absorption bands of several molecules, as indicated in the figure. Note however that two bands, namely O$_3$ (at 4.8\,$\mu$m) and HNO$_3$ (at 11.3\,$\mu$m) are only evident in transmission spectra due to the longer path through the atmosphere for this transit geometry.

We note that in emission spectra CO$_2$ features emission peaks at 4.3 and 15\,$\mu$m for the G and F-star cases (Fig. \ref{HighResSpecs}, right) due to the atmospheric temperature inversion. For a spectral resolution of $R=3\,000$ the brightness temperatures even exceed the model surface temperatures for the planets orbiting the F-type star due to the hot stratosphere (see Fig. \ref{Temps}). The spectral response seen for CO$_2$ bands is exclusively related to the temperature response since the atmospheric CO$_2$ content remains constant by construction.

The strongest spectral response on increasing the orbital distance is visible for H$_2$O, since it not only features strong absorption bands at 2.7, 6.3 and above 17\,$\mu$m, but also a strong continuum absorption over the entire wavelength range considered. On increasing the orbital distance the H$_2$O absorption clearly decreases, which is visible in both emission and transmission spectra.

The CH$_4$ bands at 3.3 and 7.7\,$\mu$m show a negligible spectral response on increasing the orbital distance (i.e. the transit depth and brightness temperature in the band remain constant), although the spectrum in the vicinity of the bands changes drastically due to the H$_2$O response. This is because the CH$_4$ absorption increases with increasing orbital distance, whereas the H$_2$O absorption decreases. For the planet around the K-type star the increase in CH$_4$ absorption even overcomes the decrease in H$_2$O absorption and both bands absorb slightly stronger for the outer runs than for the inner runs.

The response of O$_3$ (at 4.8 and 9.6\,$\mu$m) and HNO$_3$ (at 11.3\,$\mu$m) is not visible due to the strong H$_2$O response. However, a strong response was not expected from the chemical analysis. N$_2$O only features thin absorption bands at 4.5 and 7.8\,$\mu$m, which are mainly masked by other absorption bands. An absorption band of N$_2$O is visible in the vicinity of the 4.3\,$\mu$m CO$_2$ band in emission spectra, e.g. for the planets around the G and F-type star, but only in the case of a strong atmospheric temperature inversion, when the 4.3\,$\mu$m CO$_2$ band is seen in emission.

Atmospheric windows which provide a high atmospheric transmissivity provide in principle information on surface conditions. For Earth-like planets the atmosphere is transparent down to the surface at around 2.2, 3.7 and 11\,$\mu$m for high zenith angles, i.e. only for secondary eclipse observations. In primary eclipse, the atmosphere becomes opaque above about 20\,km at these wavelengths. However, the two near-IR windows cannot be used due to the reflected stellar light (see above). The brightness temperature in the 11\,$\mu$m window is equal to the surface temperature for all outer runs, whereas for the inner runs, the brightness temperature found is about 9\,K lower than the model surface temperature. This is due to the increase in optical thickness by higher water vapor concentrations.

\subsection{Filter response} \label{FilterResponse}
Photometric instruments provide high S/Ns, but at the cost of spectral information. To detect an absorption band, measurements in the filter of interest should be compared to measurements in so-called reference filters located in spectral regions where the atmosphere is assumed to be almost transparent. The measurable quantity is then the difference in the transit depth that is measured in both filters for primary eclipse transmission observations or the difference in the measured brightness temperature for secondary eclipse observations. Table \ref{Ins} summarizes the filters used within this work. The position of the filters considered in this paper are shown in Fig. \ref{HighResSpecs}.

As reference filters we will use the NIRCam F356W (3.56\,$\mu$m) and MIRI IM04 (11.3\,$\mu$m) filters. The F356W filter covers the atmospheric window around 3.7\,$\mu$m, but also the 3.3\,$\mu$m CH$_4$ absorption band. The IM04 filter covers the 11\,$\mu$m atmospheric window, but also the H$_2$O continuum absorption and in the case of transmission spectra an HNO$_3$ absorption band at 11.3\,$\mu$m.

We note that using a modified reference filter (e.g. a smaller F356W filter avoiding the CH$_4$ absorption) is beyond the scope of this paper, since we only concentrate on the filters planned for JWST.

\begin{table}
\centering
\caption{Geometric transit depth $d_\mathrm{Geo}=(r_\mathrm{P}/r_\mathrm{S})^2$ for planets orbiting different central stars.}\label{GeoTransDepth}
\begin{tabular}{cc}
  \hline  \hline
  Central star type & $d_{geo}$ \\\hline
  F-type & $7.48\times10^{-5}$ \\
  G-type & $8.4\times10^{-5}$ \\
  K-type & $1.28\times10^{-3}$ \\ \hline
\end{tabular}
\end{table}

\subsubsection{Primary eclipse}
\begin{figure*}[ht]
\centering
\resizebox{\hsize}{!}{\includegraphics*{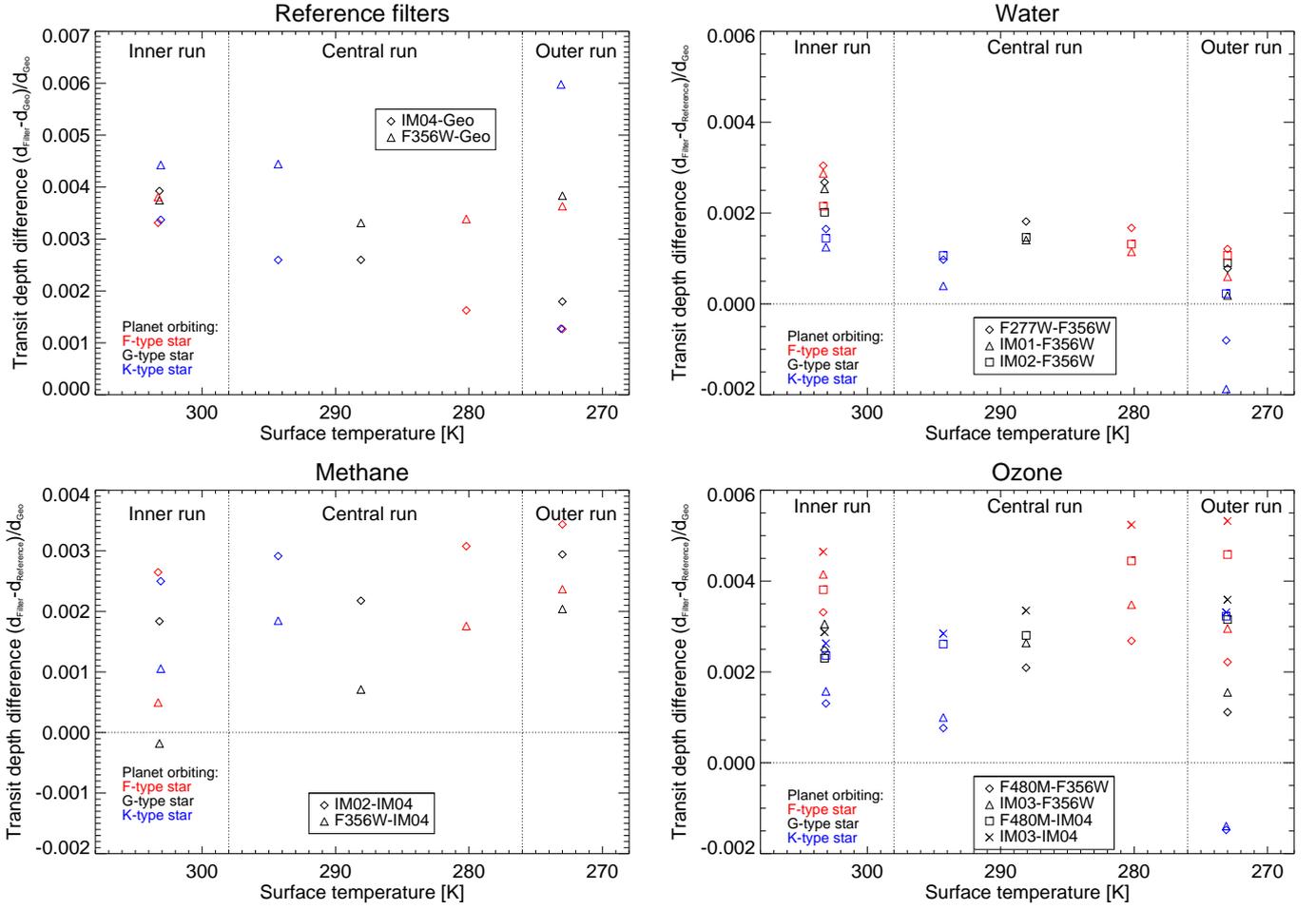}}
  \caption{Transit depth differences for different filter combinations for primary eclipse transmission spectra. $d_\mathrm{Filter}$ and $d_\mathrm{Reference}$ are the transit depths measured by the photometric filter and the reference filter, respectively, whereas $d_\mathrm{geo}$ is the geometric transit depth (i.e. the ratio of planetary and stellar radius squared).}\label{Transmissions}
  \end{figure*}

First, we compare the transit depth $d$ in the reference filters $d_\mathrm{Filter}$ with the geometric transit depth $d_\mathrm{Geo}$ given in Table \ref{GeoTransDepth}, in order to estimate the systematic error due to absorption bands covered by the reference filters. The geometric transit depth is the squared ratio of the planetary to the stellar radius. For the planetary radius $r_\mathrm{P}$ we use the Earth's radius and the stellar radii $r_\mathrm{S}$ from Table \ref{ScenarioTable}. The upper left plot in Fig. \ref{Transmissions} shows the difference of transit depth measured in the reference filters to the geometric transit depth.

For the planets around the F and G-type star, the transit depth difference $d_\mathrm{F356W}$-$d_\mathrm{geo}$ (triangles) remains about the same over the entire HZ, since the increase in the CH$_4$ abundance compensates for the decreasing scale height when increasing the orbital distance. Note that since the filter does not correctly track the decreasing atmospheric scale height with increasing orbital distance, this will lead to systematic errors when estimating the abundance of any atmospheric constituents when using this reference filter.

The CH$_4$ absorption which is covered by the filter is strongest only for the outer run of the planet around the K-type star, and thus the F356W (3.56\,$\mu$m) filter has to be used with caution as a reference and is hence not appropriate, especially when aiming at the detection of CH$_4$. A reference filter spanning from 3.5 to about 4\,$\mu$m would no longer cover the CH$_4$ absorption band and would thus serve as a better reference filter in the near-IR.

The transit depth difference $d_\mathrm{IM04}$-$d_\mathrm{Geo}$ (diamonds) decreases with increasing orbital distance due to both the decreasing atmospheric scale height and the decrease of the H$_2$O continuum absorption covered by the filter. The abundance of HNO$_3$ did not change greatly over the HZ and has thus no influence on the IM04 (11.3\,$\mu$m) filter.\\

H$_2$O absorption bands are covered by the F277W (2.77\,$\mu$m), IM01 (5.6\,$\mu$m), IM02 (7.7\,$\mu$m) and IM07 (18.0\,$\mu$m) filters (see Table \ref{Ins}). We only use the F356W (3.56\,$\mu$m) filter as a reference since the IM04 (11.3\,$\mu$m) reference filter covers the H$_2$O continuum absorption. The spectral response would thus not be visible. The strongest difference can be obtained from the F277W-F356W and IM01-F356W filter combinations (diamonds and triangles in Fig. \ref{Transmissions}, upper right plot). This is related to the decrease in H$_2$O absorption in the F277W and IM01 filter while the CH$_4$ absorption increases in the F356W reference filter, when increasing the orbital distance.

Note that the transit depth difference in the F277W-F356W and IM01-F356W filters becomes negative for the outer runs of the planets around the K-type star. In this case the CH$_4$ absorption in the F356W filter is even stronger than the H$_2$O/CO$_2$ absorption in the F277W filter and the H$_2$O absorption in the IM01 filter.

Note furthermore that the F277W (2.77\,$\mu$m) filter covers both a CO$_2$ and the H$_2$O absorption band. In order to determine whether both molecules or only one them is present in the atmosphere, H$_2$O and CO$_2$ need to be detected independently in separate absorption bands, like e.g. in IM01 (at 5.6\,$\mu$m) for H$_2$O and F430M or IM06 (at 4.3 and 15\,$\mu$m, respectively) for CO$_2$ see below.

The signal difference from the IM02-F356W filter combination (boxes) is slightly lower than for the afore-mentioned F277W-F356W and IM01-F356W combinations since the IM02 (7.7\,$\mu$m) filter also covers an CH$_4$ absorption band at 7.7\,$\mu$m, which is increasing with increasing orbital distance. However, this filter combination would help in order to minimize the effect of the CH$_4$ absorption covered by the reference filter. The IM07-F356W filter combination (at 18.0 and 3.56\,$\mu$m, respectively) features the lowest signal and is thus not shown here.\\

O$_3$ absorption bands are covered by the F480M (4.8\,$\mu$m) and IM03 (10.0\,$\mu$m) filters. The strongest signal can be inferred from the IM03-IM04 filter combination, showing a slightly increasing O$_3$ absorption with increasing orbital distance (see Fig. \ref{Transmissions}, lower right plot). This is almost exclusively an effect of the decreasing H$_2$O continuum absorption covered by the IM04 (11.3\,$\mu$m) reference filter when increasing the orbital distance, since the atmospheric O$_3$ content remains nearly constant (see Table \ref{DUs}). When using the F356W (3.56\,$\mu$m) filter as a reference also the influence of the CH$_4$ absorption in this reference filter is clearly visible. Due the strong CH$_4$ absorption in the F356W filter, the transit differences F480M-F356W and IM03-F356W become negative for the outer runs of the K-type star.\\

CH$_4$ absorption bands are covered by the F356W filter (which is also a reference filter) or by the IM02 (7.7\,$\mu$m) filter, which also covers part of the strong H$_2$O absorption band at 6.3\,$\mu$m. Thus in principle only the IM04 reference filter can be used. Both filter combinations (F356W-IM04 and IM02-IM04) clearly increase with increasing orbital distance (see Fig. \ref{Transmissions}, lower left plot) due to the increasing CH$_4$ absorption in IM02. However, as stated above also the F277W-F356W and IM01-F356W filter combinations (which were used for the detection of H$_2$O) and the F480M-F356W and IM03-F356W filter combinations (for the detection of O$_3$) can be used for the detection of CH$_4$ for certain scenarios. For the outer runs of the planets orbiting a K-type star the absorption of CH$_4$ in the F356W filter exceeds the absorption of H$_2$O/CO$_2$ in the F277W (2.7\,$\mu$m) filter, the absorption of O$_3$ in the F480M (4.8\,$\mu$m) and IM03 (10.0\,$\mu$m) filters and the absorption of H$_2$O in the IM01 (5.6\,$\mu$m) filter. The differences become negative (see diamonds and triangles in Fig. \ref{Transmissions}, upper right plot and lower right plot). \\

CO$_2$ absorption bands are covered by the F277W (2.77\,$\mu$m), F430M (4.3\,$\mu$m) and IM06 (15.0\,$\mu$m) filters. Since the CO$_2$ content in our model atmospheres remains constant, each filter combination would only show the response of the reference filter, which are also covering CH$_4$ and H$_2$O absorption bands. Nevertheless the transit depth differences would be the strongest compared to all other filter combinations, reaching a transit depth difference of up to ($d_\mathrm{F430M}$-$d_\mathrm{IM04}$)/$d_{geo}$=0.013 (not shown).\\

\subsubsection{Secondary eclipse}
\begin{figure*}[ht]
\centering
\resizebox{\hsize}{!}{\includegraphics*{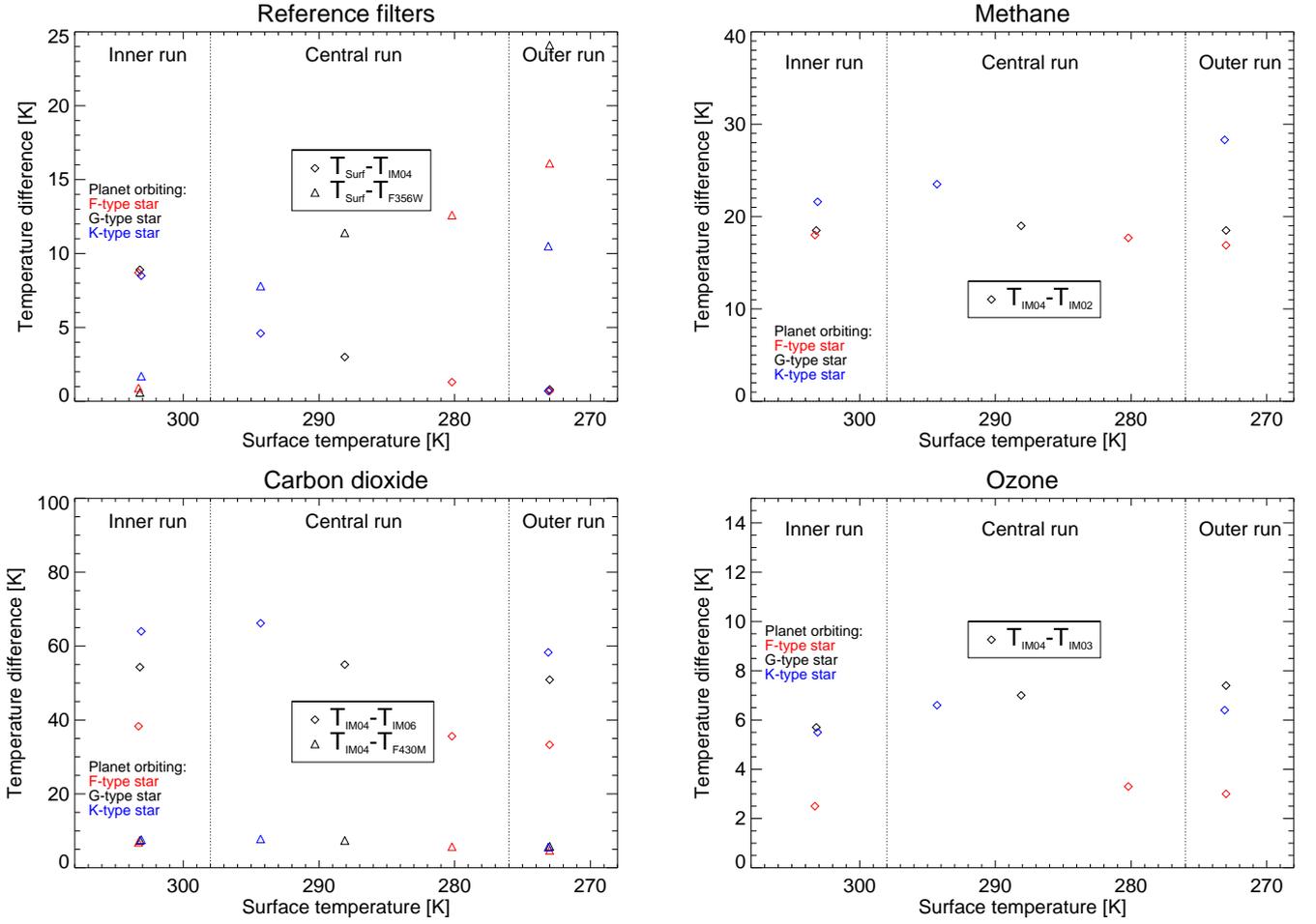}}
  \caption{Brightness temperature difference for different filter combinations for secondary eclipse emission spectra. The upper left panel shows the difference of the brightness temperature measured in the F356 and IM04 filter with respect to the model surface temperature, whereas the other panels show the difference of selected filters with respect to the reference filter IM04. Note the different vertical scales in the plots.}\label{Emissions}
 \end{figure*}

From secondary eclipse observations, the brightness temperature measured in atmospheric windows provides estimates of the temperature in the lower troposphere and potentially the surface, where the emission originates for Earth-like atmospheres. The atmospheric water vapor content controls how deep the troposphere can be probed in the atmospheric window. For example for the inner runs the broadband water continuum absorption only allows the probing of the lower troposphere and not the surface. Reflected stellar light is dominant up to 4\,$\mu$m and thus influences measurements in the near-IR filters F277W and F356W (at 2.77 and 3.56\,$\mu$m, respectively). The near-IR reference filter F356W is furthermore influenced by atmospheric CH$_4$ absorption at 3.3\,$\mu$m. This is clearly visible in Figure \ref{Emissions} (upper left plot), which shows the brightness temperature difference $T_\mathrm{Surf}$-$T_\mathrm{F356W}$, with $T_\mathrm{surf}$ being the model surface temperature and $T_\mathrm{F356W}$ the brightness temperature measured in the F356W filter. Note that the brightness temperature difference $T_\mathrm{Surf}$-$T_\mathrm{F356W}$ (triangles) increases with orbital distance since the reflected stellar light changes only slightly, whereas the surface temperature decreases by 30\,K (by construction (see Table \ref{ScenarioTable}).

If it would be possible to measure only the planetary emission, a temperature difference of about 7\,K to the model surface temperature would be found. Since also the reflected stellar light needs to be taken into account, which increases the measured signal in the near-IR, the measured temperature difference is only about 1\,K for the inner runs. For the outer runs the CH$_4$ absorption is strongest, especially for the planets around the K-type star, thus dropping the brightness temperature difference slightly. Thus, the F356W filter can not be used, neither as a reference filter nor as a filter in order to estimate surface temperatures during secondary eclipse. Therefore, we will only use the IM04 (11.3\,$\mu$m) filter as a reference filter in the following analysis.

The IM04 (11.3\,$\mu$m) reference filter is not influenced by the reflected stellar signal. It can be clearly seen that the difference $T_\mathrm{Surf}$-$T_\mathrm{IM04}$ decreases with increasing orbital distance (diamonds). For all inner runs, H$_2$O provides a strong continuum absorption over the entire IM04 filter, which influences the surface temperature determination. Inferred temperatures are thus lower than the model surface temperature by up to 9\,K. For the outer runs, which are less influenced by H$_2$O, the IM04 filter provides a surface temperature difference of below 2\,K.\\

The strongest signal for the detection of H$_2$O absorption bands can be found when comparing the IM01 (5.6\,$\mu$m) filter with the IM04 (11.3\,$\mu$m) reference filter (not shown). A brightness temperature difference of up to 40\,K for all inner runs can be found. This difference decreases with increasing orbital distance. This is because the surface temperature (visible in the reference filter) decreases stronger than the temperature in the atmospheric layers contributing to the signal measured in the measurement filter IM01, when increasing the orbital distance.\\

CH$_4$ features absorption bands at 3.3 and 7.7\,$\mu$m. Due to the reflected stellar light, we omit the 3.3\,$\mu$m band here. The 7.7\,$\mu$m band (covered by the IM02 filter) is in most cases dominated by the band wings of the broad 6.3\,$\mu$m H$_2$O band, which is clearly visible in the brightness temperature difference $T_\mathrm{IM04}$-$T_\mathrm{IM02}$ (see Fig. \ref{Emissions}, upper right plot). The temperature difference remains nearly constant for the planets around G and F-type stars, since the increase in CH$_4$ absorption compensates for the decreasing H$_2$O continuum absorption in both filters. For the planets around the K-type star however, the temperature difference is increasing due to the strong increase in CH$_4$ absorption, when increasing the orbital distance.\\

In the brightness temperature spectra shown in Fig. \ref{HighResSpecs}, CO$_2$ features strong absorption bands at 2.7\,$\mu$m, at 4.3\,$\mu$m and 15\,$\mu$m. We note that the 2.7\,$\mu$m band overlaps with an absorption band of H$_2$O and is furthermore influenced by the reflected stellar light. We thus do not take the F277W filter into account. The CO$_2$ bands probe the lower to upper stratosphere, hence are extremely sensitive to stratospheric temperatures, which differ greatly between the different central star types (see e.g. \citealt{Vasquez2013}). This is related to the lower production of O$_3$ in the atmosphere of the planet orbiting the K-type star and a weaker stellar radiation in the stellar UV bands in which O$_3$ absorbs, which both result in a colder stratosphere (see e.g. G07 and \citealt{Segura2003}). Since surface temperatures are approximately equal for all scenarios (by construction), temperature differences between the brightness temperature and the surface temperature are consequently much larger for the planets around the K-type star (cf. Fig. \ref{Temps}).

\begin{figure*}[ht]
\centering
\includegraphics[width=9cm]{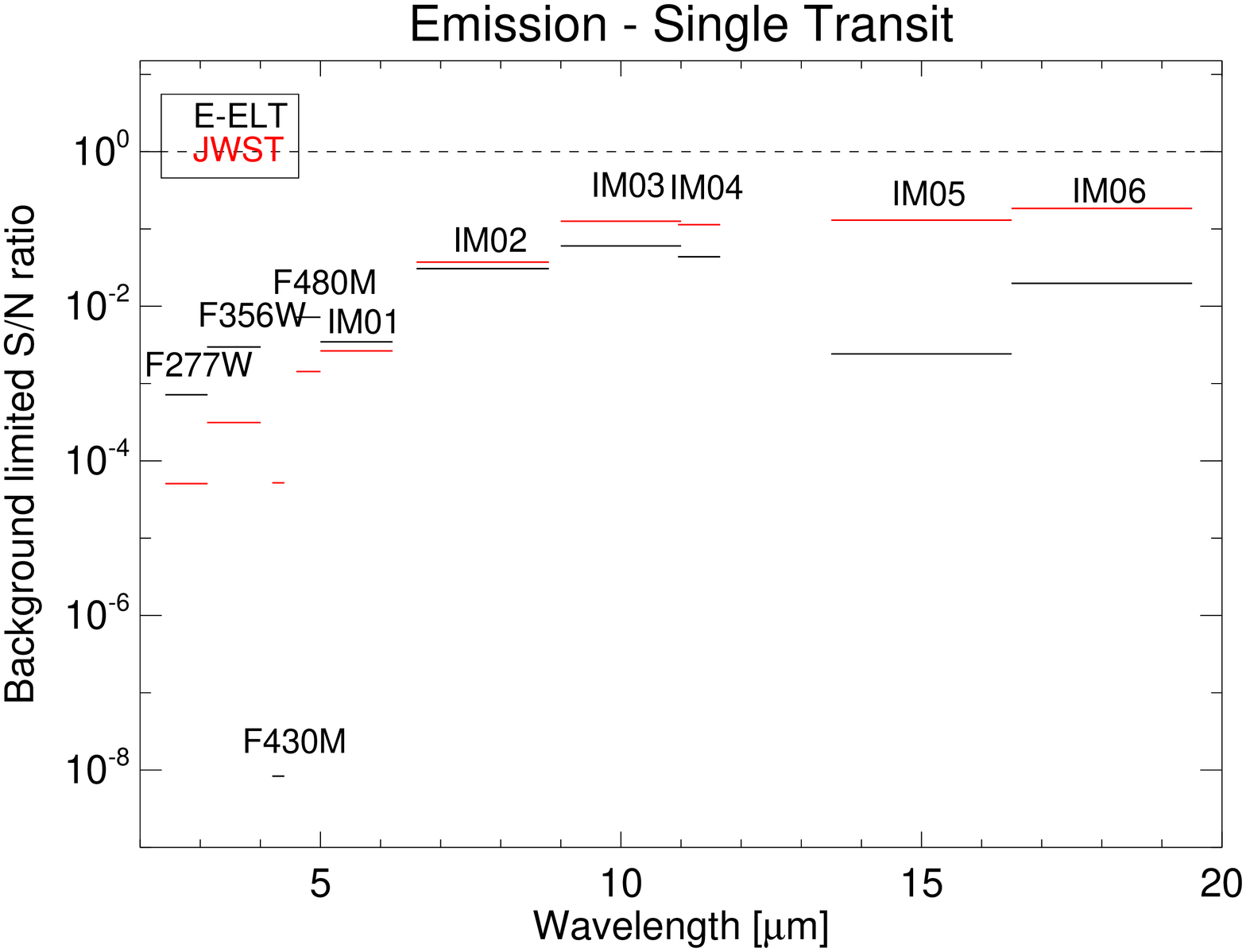}
\includegraphics[width=9cm]{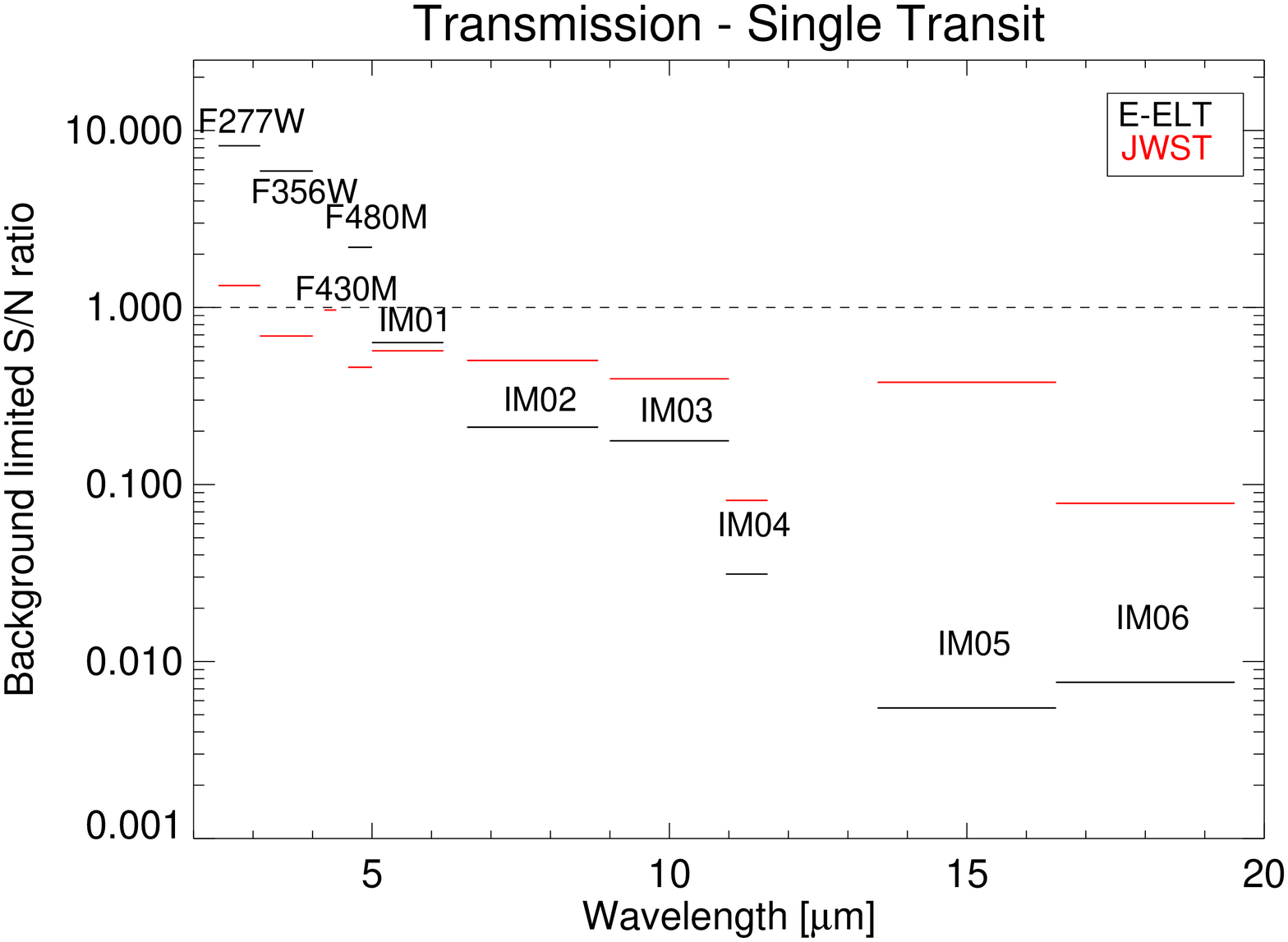}\\
\includegraphics[width=9cm]{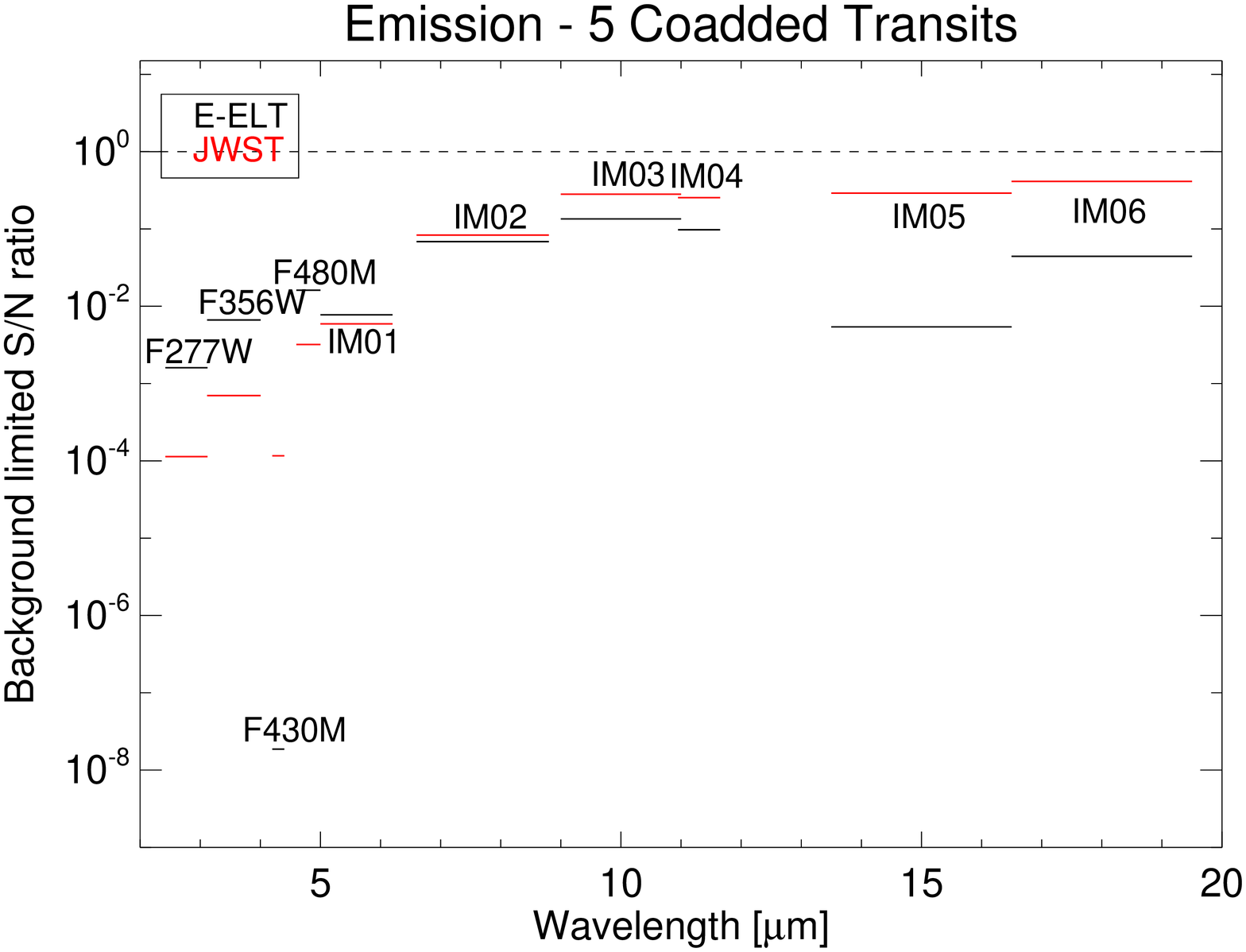}
\includegraphics[width=9cm]{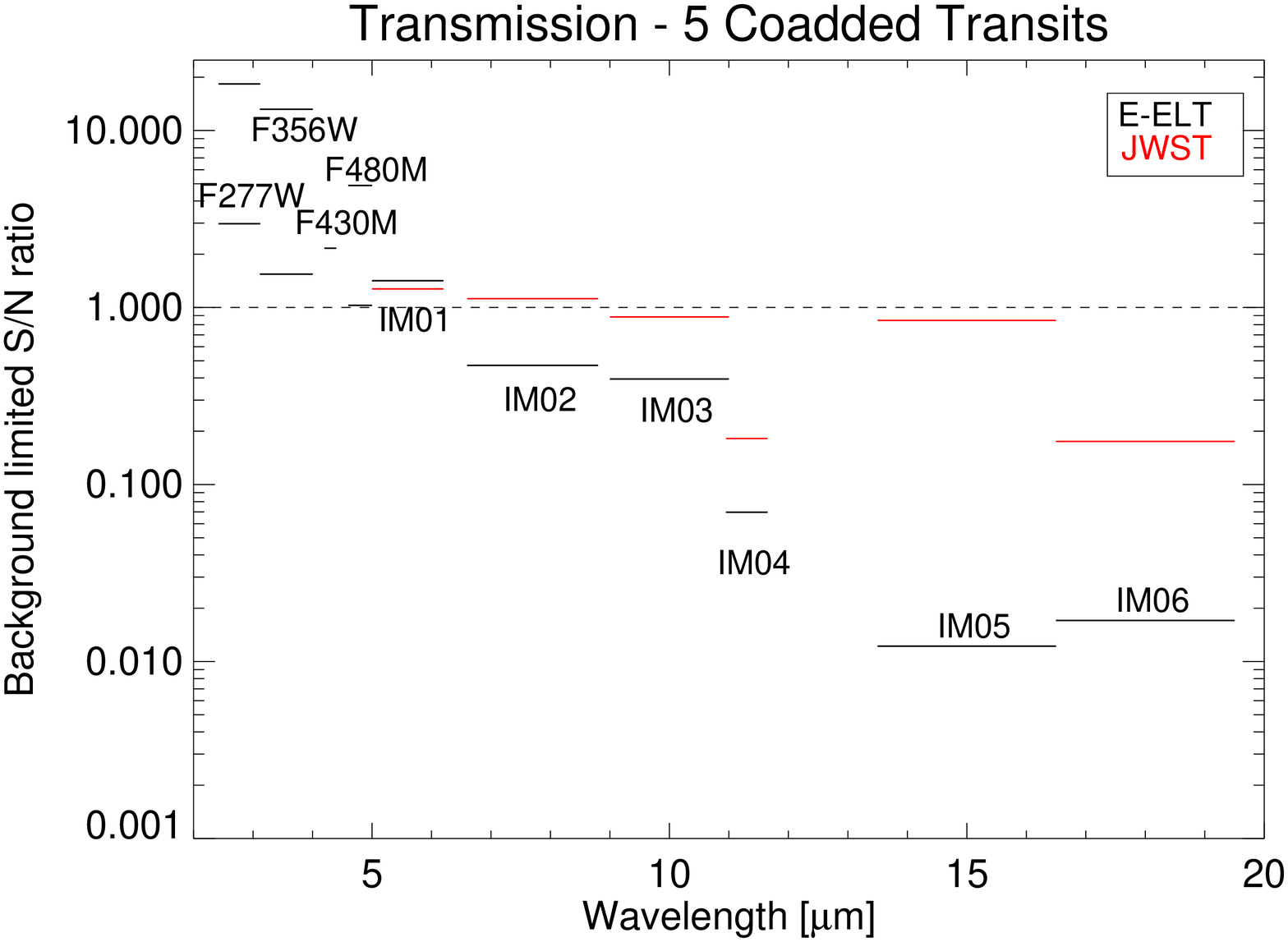}
  \caption{Background limited S/Ns for secondary (right column) and primary transit observations (left column). The S/Ns for the central run of a planet orbiting a G-type star for two different telescope configurations are shown. The upper row shows S/Ns for one single transit observation, the lower row when co-adding five observations. The solid bars represent the filter bandpasses of NIRCam from 2 to 5\,$\mu$m and of MIRI from 5 to 20\,$\mu$m. Note that the F430M filter in transmission spectra is not visible due to a low S/N below 0.001.}\label{SNRs}
  \end{figure*}

The strongest signal of up to about 70\,K can be found for the planets around the K-type stars when comparing the brightness temperatures in the IM06 (15.0\,$\mu$m) filter with the IM04 (11.3\,$\mu$m) reference filter (diamonds in Fig. \ref{Emissions}, lower left plot). Comparing the F430M (4.3\,$\mu$m) filter with the IM04 reference filter yields only modest temperature differences below 10\,K (triangles). The planets around the G and F-type stars furthermore feature lower brightness temperature differences, since the absorption in the  CO$_2$ band centers originate in the stratosphere at higher atmospheric temperatures due to the atmospheric temperature inversion (see Fig. \ref{HighResSpecs}). Thus the brightness temperature in the filter covering these bands are higher, hence the difference to the reference filter IM04 is much lower. A low brightness temperature difference would thus indicate a low-CO$_2$ atmosphere or an atmosphere with a temperature inversion. The slight decrease in brightness temperature difference with increasing orbital distance for all scenarios considered is related to the decreasing H$_2$O continuum absorption.\\

In secondary eclipse spectra O$_3$ features only an absorption band at 9.6\,$\mu$m in which the brightness temperature difference between the IM03 (10.0\,$\mu$m) filter and the IM04 (11.3\,$\mu$m) reference filter is rather low (see Fig. \ref{Emissions}, lower right plot). Since the atmospheric response on increasing the orbital distance is weak (see Table \ref{DUs}) the filter response seen is only a result of the molecular absorption bands covered by the reference channels.

\subsection{Detectability}\label{SNR}
The detectability of molecular absorption bands is investigated by calculating background-limited S/Ns for two different telescope configurations, namely JWST and E-ELT.

  \begin{figure*}[ht]
\center
\includegraphics[width=9cm]{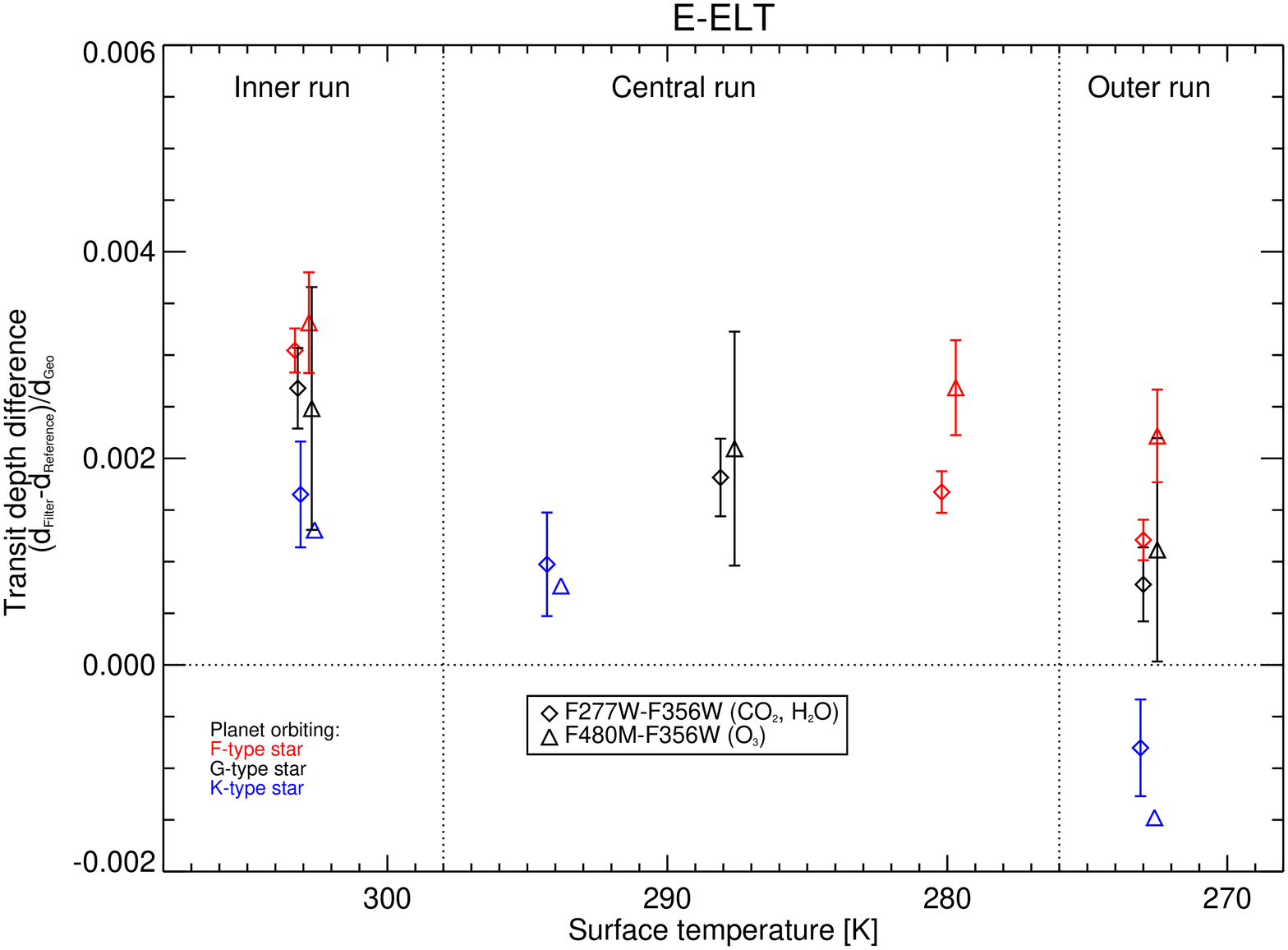}
\includegraphics[width=9cm]{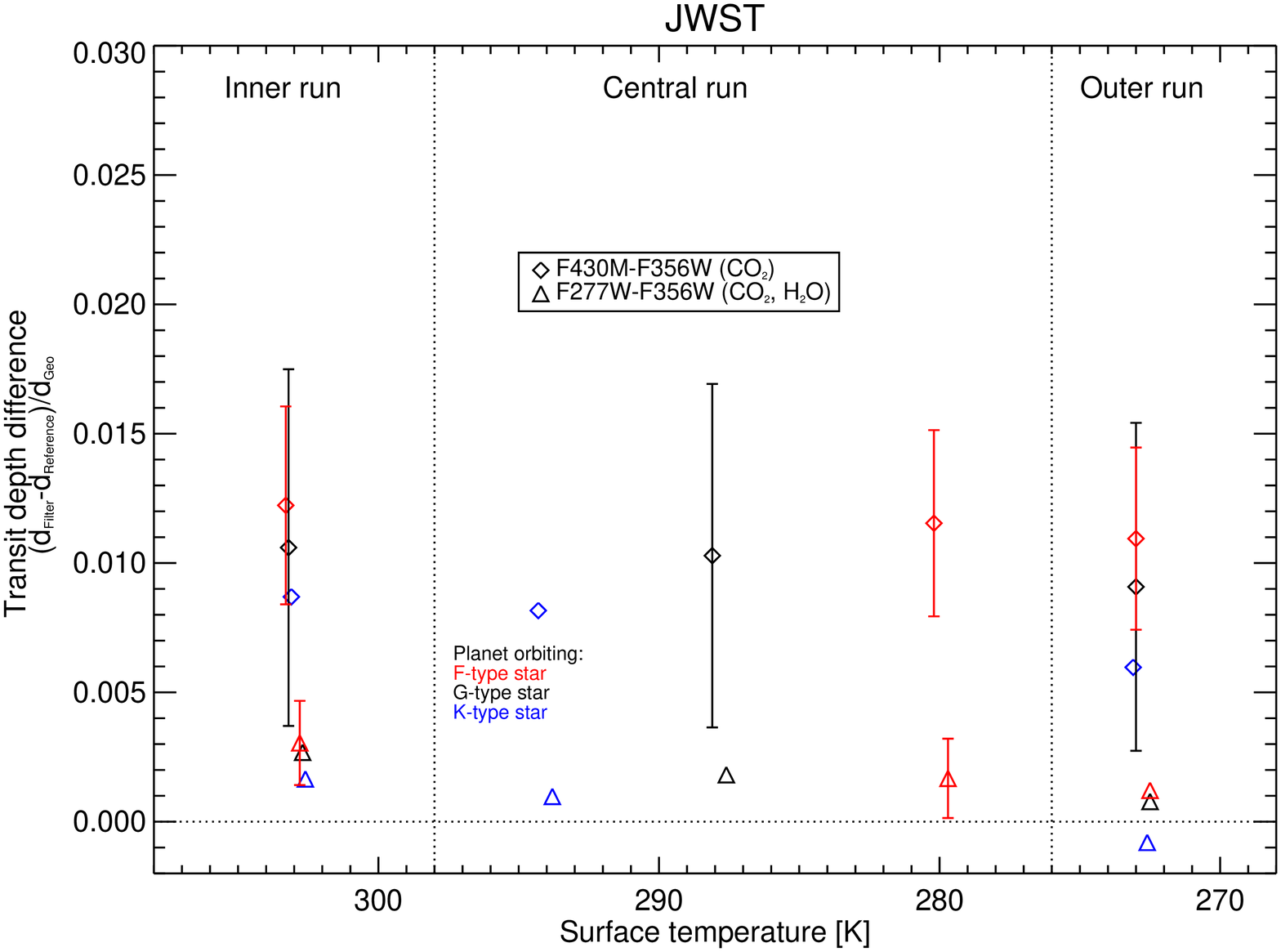}
  \caption{Transit depth differences for selected filter combinations (indicated by symbols) that provide S/Ns above unity, when co-adding five transits. Differences are shown for an E-ELT (left) and a JWST configuration (right) for the Earth-like planets around different types of central stars (F-star: red, G-star: black, K-star: blue) at different orbital distances (increasing from left to right, indicated by the increase in surface temperature). Error bars are only shown for filter combinations with differences larger than zero within their 1$\sigma$ uncertainty. For clarity reasons the different filter combinations are shifted by 0.5\,K.}\label{Filtererrors}
\end{figure*}

\subsubsection{Calculated S/Ns}
We calculate S/Ns for a hypothetical planet at a distance of 10\,pc for one single transit observation (either primary or secondary transit with transit times shown in Table \ref{ScenarioTable}). In general the background-limited S/Ns are higher for primary eclipse transmission spectra than for secondary transit emission spectra, since the absorption of the stellar light is measured instead of the much fainter emission of the planet itself. Figure \ref{SNRs} shows the S/Ns for the central run of the G-type star in the bandpass of the photometric instruments. Since the Earth's atmosphere is opaque at wavelengths where CO$_2$ absorbs, the S/Ns for ground-based telescopes are low in the band centers of the 2.7, 4.3 and 15\,$\mu$m CO$_2$ bands. Since the filters F277W and IM06 at 2.7 and 15\,$\mu$m, respectively, are broader than the absorption band, the S/Ns in these filters are dominated by the band wings.

In secondary eclipse emission spectra the highest S/Ns are obtained in the mid-IR above 10\,$\mu$m (see Fig. \ref{SNRs}, left panel) with values up to S/N=0.3 during a single transit observation. Note that ground-based telescopes are limited by the thermal emission of the telescope and Earth's atmosphere. By contrast, primary eclipse transmission observations provide the highest S/Ns in the near-IR (see Fig. \ref{SNRs}, right panel), with S/Ns of up to 9 during a single transit observation for the near-IR filters for ground-based observations with the E-ELT. At wavelengths longer than 6\,$\mu$m the thermal emission of the E-ELT dominates the overall noise.\\

We note that higher S/Ns may be conceivable when observing bigger planets (for secondary transit emission spectroscopy) around stars that are closer to the observer (the S/N increases quadratically with the planetary radius and decreases linearly with the distance, see e.g. \citealt{Rauer2011}). The distance to the planet is especially important for wavelengths where the noise is photon-dominated (i.e. in the near-IR). The zodiacal noise source decreases for targets at higher ecliptic latitudes. Currently we use measurements in the ecliptic plane for our zodiacal noise \citep{Kelsall1998}, which is a pessimistic assumption. This will be particularly important for measurements in the thermal-IR (i.e. in the MIRI IM06 filter).  In addition, co-adding of transits could lead to increased S/Ns if systematic effects can be controlled. For example five transits can in principle be observed for a planet around a G-type star within the projected five-year mission time of JWST (i.e. the S/Ns increase by a factor of about two, see bottom row of Fig \ref{SNRs}). Finally the telescope and instrument designs can be improved to obtain higher S/Ns. The S/N increases linearly with the telescope's aperture and with the square-root of the quantum efficiency (see, e.g. \citealt{Rauer2011}). Also the thermal emission of a space-borne telescope might be lowered. In total, a factor 10 higher S/Ns are conceivable.

\subsubsection{Detectability of molecular absorption bands using filters}
From the calculations shown above it is clear that the S/Ns are too low for a single transit observation in order to detect any absorption band in secondary eclipse spectra. For primary eclipse observations in the near infrared three filters provide S/Ns of larger than unity, when using an E-ELT-like telescope configuration (namely F277W at 2.7\,$\mu$m, F356W at 3.56\,$\mu$m, and F480M at 4.8\,$\mu$m). These filters are covering the H$_2$O/CO$_2$ band at 2.7\,$\mu$m (F277W), an atmospheric window at 3.7\,$\mu$m which can be used as a reference (F356W) and an O$_3$ band at 4.8\,$\mu$m (F480M). When using a JWST-like telescope configuration, only the F277W and F430M (4.3\,$\mu$m) filter provide S/Ns above unity during a single transit observation. The latter covers a CO$_2$ absorption band at 4.3\,$\mu$m.

During a single primary transit observation with the ELT (not shown), an absorption band in the F277W and F480M filter can be detected only for the planets around the F-type star. For planets orbiting a G-type star only an absorption band in the F277W filter can be detected, whereas no absorption bands are detectable for planets around K-type stars. Using the JWST, the CO$_2$ absorption band at 4.3\,$\mu$m can be detected in the F430M filter, but only for the planets around the F-type star. This is due to the strong transit depth difference of up to 0.013 between the F430M and the F356W reference filter. The S/N for the other stellar types are too low for a single transit observation.

Since both CO$_2$ and H$_2$O feature an absorption band at 2.7\,$\mu$m, combining observations of JWST and E-ELT it is thus possible to determine the presence of H$_2$O in the atmosphere of the planet around an F-type star: If the presence of a CO$_2$ absorption at 4.3\,$\mu$m can be ruled out using JWST measurements and an absorption band can be detected at 2.7\,$\mu$m using E-ELT, than this might be attributed to the presence of H$_2$O in the atmosphere. However, in order to detect H$_2$O in a separate band (e.g. at 6.3\,$\mu$m), a S/N$>7$ needs to be achieved, hence about 10 transits need to be co-added for an E-ELT-telescope setup.

We assume here that during the five-year mission time of JWST about five transits of an Earth-like planet orbiting a main-sequence star in the HZ can be observed, assuming that instrumental effects can be neglected. This would increase the S/N by a factor of about two. Figure \ref{Filtererrors} shows the transit depth differences in the above-mentioned filters with 1$\sigma$ error bars for five transit observations. The E-ELT configuration (left plot) would now allow the detection of both the 2.7\,$\mu$m CO$_2$/H$_2$O band and the 4.8\,$\mu$m O$_3$ band in the F277W (2.77\,$\mu$m) and F480M (4.8\,$\mu$m) filters, respectively, for planets around G and F-type stars. For planets orbiting K-type stars only the 2.7\,$\mu$m band can be detected. Most interestingly, a negative transit depth difference between the F277W and F356W filter in the outer run of the K-type planets can be detected. This is due to the strong 3.3\,$\mu$m CH$_4$ absorption covered by the F356W filter. Hence it would be possible to prove the existence of CH$_4$ in the planetary atmosphere. Using the JWST (right plot) now also allows the detection of the 4.3\,$\mu$m CO$_2$ band for planets around G-type stars. Furthermore, the 2.7\,$\mu$m CO$_2$/H$_2$O band can be detected for planets around F-type stars.

\section{Discussion}\label{Discussion}
We note that clouds can have a significant impact on the spectral appearance of a planet, as has been investigated by \citet{Robinson2011} and \citet{Kitzmann2011}. With increasing cloud cover of either low or high-level clouds, \citet{Kitzmann2011} found that the overall IR emission of an Earth-like planet decreases and the absorption bands of O$_3$, CO$_2$ and H$_2$O are weakened. Hence our S/Ns for emission spectra are already likely to be upper limits. For this work this means that the brightness temperature difference between the filter covering the 9.6\,$\mu$m O$_3$ band and the reference filter centered at 11.3\,$\mu$m will decrease, when taking into account clouds. Furthermore, the uncertainty in the surface temperature determination will increase.

For primary eclipse transmission spectra, low level clouds have only a negligible effect, since the cloud layers are present below the tangential height, where the atmosphere is already optically thick. However, in order to study the effect of mid- or high-level clouds, 3-dimensional atmospheric and radiative transfer models need to be applied in order to account correctly for the absorption and scattering by the cloud layers.\\

The main challenge for ground-based observations of an Earth-like planet is to remove the contamination due to Earth's atmospheric absorption. High-resolution observations may take advantage of the Doppler shift of single lines due to the proper motions of the star, the exoplanet, the Sun, and the Earth (see e.g. \citealt{Vidal-Madjar2010}), to distinguish between telluric absorption lines and lines forming in the atmosphere of the exoplanet. However, this is not applicable to absorption bands for low-resolution spectroscopy or photometry. These observations, however, may use multi-object spectroscopy or photometry in order to observe nearby reference stars, as in \citet{Bean2010}.\\

We were able to show that the 4.8\,$\mu$m absorption band of the biomarker molecule O$_3$ and the absorption band at 6.3\,$\mu$m of the related compound H$_2$O are both detectable during primary eclipse, when co-adding several transit observations. The former molecule is on Earth directly related to the presence of plants and cyanobacteria that produce O$_2$ from photosynthesis, whereas the latter is needed in its liquid phase on the surface for life as we know it.

Note that the detection of the presence of O$_3$ in an exoplanetary atmosphere does not necessarily indicate the presence of a biosphere. It is currently discussed, if O$_3$ can be formed in an abiotic way by the photodissociation of CO$_2$ in e.g. a CO$_2$ dominated atmosphere \citep{Selsis2002} or in an atmosphere which is subject to strong UV irradiation (e.g. around active M-dwarfs). A combination of strong UV-C and weak UV-B irradiation is required for the latter process (see \citealt{Domagal2010}). \citet{Segura2007} however showed that it is not possible to create O$_3$ from the photolysis of CO$_2$ in an Earth-like planet with an active hydrological cycle orbiting a Sun-like star.

Another way of forming substantial amounts of abiotic O$_3$ is possible in an atmosphere that is subject to strong atmospheric escape, where atmospheric H$_2$O is photolyzed, and H is efficiently removed. The atmosphere then becomes enriched in atomic oxygen, which can form O$_2$ and O$_3$ \citep{Schindler2000}. Such possible scenarios should be carefully evaluated before claiming the detection of life.

\section{Summary \& Conclusions}\label{Summary}
In this paper we have studied the spectral appearance of cloud-free Earth-like atmospheres at different orbital distances within the HZ of three different main-sequence central stars. We have investigated under which conditions molecular absorption bands are detectable with near-future instruments and telescopes, either from ground or from space, and if surface temperatures can be inferred. For this we have calculated background-limited S/Ns for an E-ELT-type ground-based telescope and a JWST-like space-borne telescope, assuming that our target planets are located at a distance of 10\,pc. For the instruments considered in this work, we used the specifications and bandpasses of two photometric instruments that are planned for the JWST. Of special interests are atmospheric species which are related to habitability (CO$_2$, H$_2$O) or could even be linked to biological activity (O$_3$, CH$_4$). Also the knowledge of surface temperature is central to the habitability problem.

In general, for an habitable Earth-sized planet, the contrast between the stellar emission and the planetary emission is too low as to provide a sufficient S/N in order to detect any absorption bands during secondary eclipse. However, during primary eclipse transmission spectroscopy much higher S/Ns (especially in the near-IR) can be obtained since the absorption of the stellar light in the planetary atmosphere is measured. Furthermore the geometry of the primary transit allows for weak absorption bands to produce significant features due to the longer lightpath in the atmosphere, which is e.g. the case for O$_3$ at 4.8\,$\mu$m and HNO$_3$ at 11.3\,$\mu$m, which are not visible in secondary eclipse emission spectra. This enhanced path length allows the detection of key compounds in an exoplanet atmosphere that are also present in the Earth's atmosphere, as long as these compounds are observed from the ground in their optically thin bands. During a lunar eclipse e.g. \citet{Palle2011} was able to detect from ground-based measurements a number of biomarker molecules as well as the dimers O$_2$-O$_2$ and O$_2$-N$_2$ in the transmission spectrum of our Earth's atmosphere.

We found that ground-based observations using the E-ELT enables the detection of near-IR absorption bands of both CO$_2$ and H$_2$O at 2.7 and O$_3$ at 4.8\,$\mu$m for some cases, even for a single transit observation. In order to discriminate whether both CO$_2$ and H$_2$O or only one of them is present, separate bands need to be observed. This is e.g. possible using a space-borne telescope like the JWST, which allows the detection of the CO$_2$ absorption band at 4.3\,$\mu$m, which is not detectable for ground-based telescopes due to absorption in the Earth atmosphere. However, in order to detect H$_2$O in a separate band at least a S/N$>7$ need to be achieved for E-ELT observations, hence about 10 transits need to be co-added.

In order to increase the S/N, the observation of planets in the HZ around M-type stars would be a solution due to a much shorter period of about 32 days for planets in the HZ. Although the integration time during the transit is much shorter (3\,h compared to 13\,h) and the stellar luminosity is much lower for M-dwarfs than for e.g. G-type stars, a factor of about 11 more transits can be observed within one year, which would translate into a factor of about three times higher S/Ns during a one year mission time. The shorter integration time and lower stellar luminosity results in a factor of about two lower S/Ns for an M-dwarf planet than that for a planet orbiting a G-type star (see e.g. \citealt{Rauer2011} for photon-limited S/Ns).\\

This paper shows for the first time, that photometric filters planned for the JWST or the E-ELT can also be used for the characterization of Earth-like exoplanet atmospheres, although the filter bandpasses have been defined for different scientific purposes. They can be used to characterize a given planetary atmosphere even if they provide only a very low spectral resolution. However, several filters need to be positioned over a broad wavelength range in order to obtain information about different biomarker molecules and surface conditions. In order to obtain information about the surface conditions and to perform a comparative filter analysis as performed in this paper, reference filters need to be chosen carefully in order not to be contaminated by spectral absorption bands. For primary eclipse transmission observations, two near-IR atmospheric windows from 2.1 to 2.4\,$\mu$m  and from 3.5 to 4.0\,$\mu$m as well as in the mid-IR at around 11\,$\mu$m could be used for reference filters when aiming at Earth-like atmospheres. The NIRCam F356W filter that has been used in this paper (which is planned for the JWST), is too broad (3.1 - 4.0\,$\mu$m) and thus cover an CH$_4$ absorption band at 3.3\,$\mu$m. The near-IR windows provide much higher S/Ns than the mid-IR window. For secondary eclipse observations only the 11\,$\mu$m atmospheric window can be used as a reference filter, since both near-IR windows feature very low S/Ns and are furthermore influenced by reflected stellar light.

\section*{Acknowledgements}
We thank the referee V. Meadows for a thorough reading of and detailed comments on the manuscript. This research has been partly supported by the Helmholtz Gemeinschaft (HGF) through the HGF research alliance "Planetary Evolution and Life". Pascal Hedelt, Philip von Paris and Franck Selsis acknowledge support from the European Research Council (Starting Grant 209622: E$_3$ARTHs). Discussions with B. Stracke and A. Belu are gratefully acknowledged.

\bibliographystyle{aa}
\bibliography{References_HZn}

\end{document}